\documentclass[%
 reprint,
 amsmath,amssymb,
 aps,superscriptaddress,
]{revtex4-2}

\usepackage{graphicx}
\usepackage{dcolumn}
\usepackage{bm}
\usepackage{hyperref}
\usepackage[mathlines]{lineno}
\usepackage{amsmath}
\usepackage{color}
\usepackage{caption}
\usepackage{subcaption}

\begin{document}

\preprint{APS/123-QED}

\title{Stabilization of Spatiotemporal Dissipative Solitons in Multimode Fiber Lasers by External Phase Modulation}

\author{V. L. Kalashnikov $^{1,2}$ and S. Wabnitz }

\address{
Dipartimento di Ingegneria dell’Informazione, Elettronica e Telecomunicazioni, Sapienza Universit\`a di Roma, via Eudossiana 18, 00184 Rome, Italy \\
$^{2}$ Department of Physics, Norwegian University of Science and Technology, H{\o}gskoleringen 5, Realfagbygget, NO-7491, Trondheim, Norway}

\email{vladimir.kalashnikov@ntnu.no}

\date{\today}

\begin{abstract}
\noindent In this work, we introduce a method for stabilizing spatiotemporal solitons. These solitons correspond to light bullets in multimode optical fiber lasers, energy-scalable waveguide oscillators and amplifiers, localized coherent patterns in Bose-Einstein condensates, etc. We show that a three-dimensional confinement potential, formed by a spatial transverse (radial) parabolic graded refractive index and dissipation profile, combined with quadratic temporal phase modulation, may permit the generation of stable spatiotemporal dissipative solitons. This corresponds to combining phase mode-locking with the distributed Kerr-lens mode-locking. Our study of the soliton characteristics and stability is based on analytical and numerical solutions of the generalized dissipative Gross-Pitaevskii equation. This approach could lead to higher energy (or condensate mass) harvesting in coherent spatio-temporal beam structures formed in multimode fiber lasers, waveguide oscillators, and weakly-dissipative Bose-Einstein condensates.

\end{abstract}

\maketitle


\section{\label{sec:intro}Introduction}

Research progress in mastering stable multidimensional wave patterns \cite{Cross,LUGIATO,WEISS} has an interdisciplinary character. It bridges across different phenomena in both physical and social sciences, ranging from photonics (so-called light bullets, LBs, or spatiotemporal, ST, optical solitons) \cite{Silberberg,Malomed1,Wise1}, Bose-Einstein condensates (BECs) \cite{Malomed2,kartashov}, plasma confinement \cite{Robinson}, ``living matter'' \cite{marchetti}, socio-technical systems \cite{vespignani}, and many other fields \cite{abdullaev,akhmediev2008dissipative}. In photonics and BEC, multidimensional soliton-like structures could provide unprecedented energy (or mass) condensation \cite{wright,our1,our2}, as well as breakthroughs in the information capacity of photonic networks \cite{richardson}, the mastering of multimode microresonators for integrated waveguide lasers, optical comb generation, optomechanics \cite{del,kippenberg,mackenzie2007dielectric}, and the generation of dissipative cavity solitons in coherently-driven Kerr resonators \cite{leo2010temporal,driven}.

The main issue hampering the practical use of multi-dimensional soliton-like structures is given by their intrinsic instability \cite{kelley}, which leads to the formation of filaments, rogue-wave-like and turbulent structures \cite{dudley2014instabilities,turitsyna2013laminar}. This challenge becomes crucial when the number of transverse and longitudinal modes in multimode fibers (MMFs) grows larger, owing to the loss of coherence between modes, and the resulting unprecedentedly complex mode-beating dynamics  \cite{turitsyna2013laminar,krupa2019multimode,wu2019thermodynamic}. Several methods for addressing multi-dimensional soliton instabilities have been recently proposed. Basic approaches involve the use of ``trapping potentials.'' In photonics, for instance, graded refractive index, i.e., GRIN or photonic crystal fibers, as well as arrays of waveguides can be used \cite{kartashov,malomed2016multidimensional,shtyrina2018coexistence,minaldi2010arrays}. In combination with Kerr nonlinearity (or attractive boson interaction in BEC), such potentials could provide transverse-mode stabilization, and even spatiotemporal soliton or LB formation \cite{krupa2017spatial,renninger2013optical}.

However, harvesting energy (or mass for BEC) involves interacting with an environment (i.e., a ``basin''). Such systems must be dissipative, and stable emerging soliton-like structures should belong to the class of the so-called dissipative solitons (DSs) \cite{akhmediev2008dissipative}, or multidimensional dissipative LB, which, in particular, could be generated in a driven Kerr-cavity \cite{tlidi2021cavities}.
As it was earlier pointed out, dissipative nonlinearities can stabilize both spatial and ST DSs \cite{grelu2005light,skarka2006stability,leblond2009stable,mihalache,malomed2014spatial}. Such stabilization mechanisms could be enhanced, in particular, by gain localization \cite{bhutta2002spatial,lam2009spatial,lobanov2012topological,malomed2014spatial,skarka2010varieties}. Nevertheless, until now an endeavor for demonstrating true ST 3D-DSs remains challenging, because it requires using some additional mode-locking (i.e., dissipative nonlinearity) mechanisms for its operation (e.g., see \cite{grelu2005light,teugin2019spatiotemporal}).

An alternative mechanism has been proposed for multi-dimensional soliton stabilization: it is based on nonlinear transverse mode coupling in fiber arrays \cite{winful1992passive,aceves1994multidimensional,proctor2005nonlinear} or tapered multicore fibers \cite{buttner2012multicore}. Such a mechanism requires changing a basic paradigm: nonlinearity (e.g., Kerr-nonlinearity in photonics or attractive boson interaction in BECs) must be enhanced rather than suppressed.
Such a change of approach (see \cite{renninger2014spatiotemporal}) was implemented by means of the so-called distributed Kerr-lens mode-locking (DKLM) technique. DKLM allows for the effective energy harvesting of femtosecond pulses in thin-disk solid-state oscillators, operating in either normal or anomalous chromatic dispersion regimes \cite{Zhang:15}. The obvious resource for energy harvesting is the up-scaling of the laser mode size, which is prone to introduce multimode instabilities. Therefore, the soliton stabilization mechanism based on increasing the level of nonlinearity was called ``ST mode-locking'' (STML), ultimately leading to LB or ST soliton generation \cite{renninger2014spatiotemporal}. This is a promising technique for the applications of MMF lasers \cite{our1} and mid-infrared waveguide lasers \cite{Sorokin:22}.

\begin{figure}[htpb]
\includegraphics[width=8.5cm]{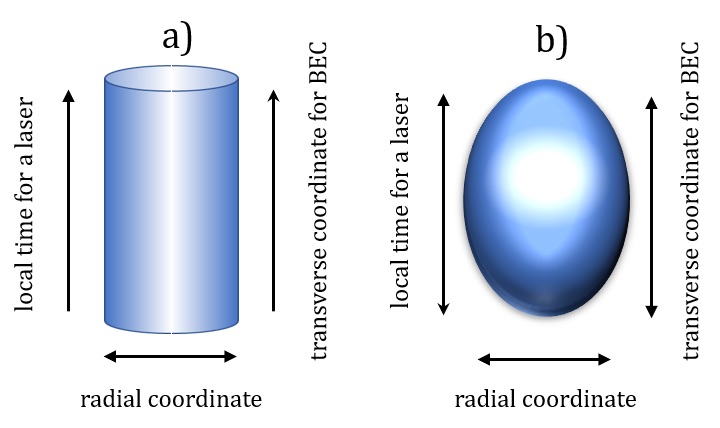}
\caption{\label{fig1} Schematic illustration of complex confining potentials of either cigar- (a) or pancake-like (b) shape in GRIN MMFs or BEC. Here the transverse coordinate $z$ in the ($z$, $r$)- frame for BECs is analogous to local time $t$ in the ($t$, $r$)- frame for a laser operating in the anomalous dispersion regime.}
\end{figure}

As it was recently shown in \cite{our1}, the concept of STML could be implemented as DKLM in a GRIN MMF laser, with transverse grading of both refractive index and dissipation. The complex confinement potential in \cite{our1} corresponds to a cigar-like confining potential in a weakly-dissipative BEC (Fig. \ref{fig1}, a) \cite{our2}.
As it was previously found, an external periodical phase modulation can substantially enhance 1D-soliton stability in a dissipative laser system in the presence of Kerr-nonlinearity \cite{Doerr:94,SMITH1993324,kalashnikov1998phase,iet:/content/journals/10.1049/el_19931138,Chang:09}. The implementation of this idea is based on using a so-called ``space-time duality'' when quadratic phase modulation plays a role of a ``time ($t-$) lens'' \cite{kolner1994space}. As a rule, a periodic phase modulation ($\propto t^2 + H.O.T.$) could be produced by an acousto- (or electro) optic phase modulator \cite{leo2010temporal,driven} (see Section E). A recirculating pulse sees an effective, path-averaged phase modulation whenever the modulator is introduced along the cavity's longitudinal coordinate, and its frequency is a harmonic of the cavity mode spacing. Such a modulation acts as time- ($t-$) dependent guiding potential, which can be approximated by a parabola around the center of the pulse.
The combination of transverse spatial and temporal harmonic potentials leads to a 3D, pancake-like potential which confines LBs in both spatial and temporal dimensions (Fig. \ref{fig1}, b), thus facilitating total mode-locking, or simultaneous transverse and longitudinal mode-locking \cite{wright}. 

In this work, we theoretically study how the presence of a 3D potential affects ST soliton stability. We optimize other dissipative parameters and demonstrate that temporal phase modulation enhances LB stability. We show that, on the one hand, periodic phase modulation may even suppress LB formation; on the other hand, it may permit to support the stable LBs, without the need for a major contribution of graded dissipation.

\section{\label{sec:analytics}Methods}

\subsection{Generalized dissipative Gross-Pitaevskii equation}

Well-established approaches use the Gross-Pitaevskii equation (GPE) for modeling the evolution of either the electric field in GRIN MMFs, or matter dynamics of BECs, in the presence of a confining external potential \cite{Malomed1,kartashov,krupa2019multimode,renninger2013optical}. We build here on a generalized GPE model, taking into account the presence of both dissipative effects \cite{our1,our2} and of a 3D or pancake-like confining potential (Fig. \ref{fig1}, b):

\begin{gather} \label{eq:1}
  i\frac{{\partial a}}{{\partial \xi }} =  - \frac{1}{2}\left[ {\frac{1}{r}\frac{\partial }{{\partial r}}r\frac{\partial }{{\partial r}} + \left( {1 - 2 i\tau } \right)\frac{{{\partial ^2}}}{{\partial {\chi ^2}}}} \right]a +\\+ \frac{1}{2}\left( {1 - 2i\kappa } \right){r^2}a + \nu {\chi ^2}a - {\left| a \right|^2}a - i\Lambda a. \nonumber
\end{gather}

\noindent Here, $\xi$ stands for either time $T$ for BECs, or the propagation coordinate $Z$ for an MMF laser, respectively. Moreover, $r$ is a radial coordinate under the condition of axial symmetry; the first term in square brackets describes diffraction in photonics or the contribution of the transverse component of boson kinetic energy in BECs. $\chi$ is a local time coordinate $t$ in a co-moving coordinate frame in photonics or a longitudinal spatial coordinate $z$ in BECs. Then, the ${{{\partial ^2}} \mathord{\left/
 {\vphantom {{{\partial ^2}} {\partial {\chi ^2}}}} \right.
 \kern-\nulldelimiterspace} {\partial {\chi ^2}}}$-term describes anomalous group-velocity dispersion in photonics or the contribution of the longitudinal component of boson kinetic energy in BECs. The pancake-like (i.e., $r, \chi$-dependent) parabolic confining potential is provided by the GRIN structure and by temporal phase modulation (parameter $\nu$) produced by, e.g., an intracavity acousto- (or electro) optic modulator. A transversely graded dissipation is defined by the $\kappa$-parameter, leading to effective on axis ($r=0$) gain ($\Lambda<0$). For BECs, a ``gain'' means an inflow from a non-coherent ``reservoir'' \cite{our2,bludov2010nonlinear}. We assume no dissipation confinement along the $t/z$-axis (i.e., no a short ($t$)-scale gain localization in a laser, or a longitudinal (i.e., structured along the $z-$coordinate) confining lattice in BECs \cite{bludov2010nonlinear,fedichev1996influence}). The nonlinear self-interaction term ${\left| a \right|^2}a$ is defined by the self-phase modulation (SPM) strength in photonics or the two-body scattering length (``attracting'') in BECs. The spectral dissipation in a laser (or ``kinetic cooling'' along the dissipative unconfined $z-$axis in BEC) is defined by a $\tau-$parameter.

 Eq. (\ref{eq:1}) is dimensionless, and the normalization rules for photonics correspond to those in \cite{our1,raghavan2000spatiotemporal}: the transverse spatial coordinate is normalized to ${w_0} = {1 \mathord{\left/
 {\vphantom {1 {\sqrt[4]{{2{k_0}\left| {{n_1}} \right|{\beta _0}}}}}} \right.
 \kern-\nulldelimiterspace} {\sqrt[4]{{2{k_0}\left| {{n_1}} \right|{\beta _0}}}}}$, where $k_0=\omega_0/c$, ${\beta _0} = n\left( {{\omega _0}} \right){k_0}$ ($\omega_0$ is a carrier frequency, $c$ is the light speed in vacuum, $n(\omega_0)$ is a refractive index on the axis $r=0$, and $\left| {{n_1}} \right|$ is a measure of refractive index change along the $r-$direction. The propagation length is normalized to $L_d={\beta _0}w_0^2$, and the co-moving frame is defined as $t = {{\left( {T - {\beta _1}Z} \right)} \mathord{\left/
 {\vphantom {{\left( {T - {\beta _1}Z} \right)} {{T_0}}}} \right.
 \kern-\nulldelimiterspace} {{T_0}}}$ ($\beta_1$ is a group-velocity coefficient). Local time $t$ is normalized to ${T_0} = \sqrt {\left| {{\beta _2}} \right| L_d}$, where $\beta_2$ is a group-velocity dispersion coefficient. The field amplitude normalization scale is $\sqrt {{k_0}{n_2}{L_d}}$, where $n_2$ is a Kerr nonlinear coefficient.

\subsection{Variational approximation}

One may conjecture that the desired result of multi-mode synthesizing is close to a fundamental mode \cite{krupa2017spatial}. In this case, the powerful variational technique could be used \cite{our1,our2,yu1995spatio,raghavan2000spatiotemporal}. The corresponding ansatz for a LB without vorticity charge can be written as:

\begin{gather} \label{eq:2}
    a\left( {Z,r,t} \right) = \alpha \left( Z \right)\,\exp \left[ {i\left( {\phi \left( Z \right) + \psi \left( Z \right){t^2} + \theta \left( Z \right){r^2}} \right)} \right] \times \nonumber\\
    \times {\mathop{\rm sech}\nolimits} \left( {\frac{t}{{\upsilon \left( Z \right)}}} \right)\exp \left[ { - \frac{{{r^2}}}{{2\rho {{\left( Z \right)}^2}}}} \right],
\end{gather}

\noindent where the propagation distance or $Z-$dependent parameters are defined as follows: $\alpha$ is an LB amplitude, $\phi$ is a phase, $\psi$ and $\theta$ are the temporal and spatial chirps, respectively, $\upsilon$ is an LB temporal width, and $\rho$ is a beam size.

The generating Lagrangian for the non-dissipative part of Eq. (\ref{eq:1}) is

\begin{gather} \label{eq:3}
    L = \frac{i}{2}\left[ {{a^*}{\partial _Z}a - a\,{\partial _Z}{a^*}} \right] + \frac{1}{2}\left( {{{\left| {{\partial _t}a} \right|}^2} + {{\left| {{\partial _x}a} \right|}^2} + {{\left| {{\partial _y}a} \right|}^2}} \right) \nonumber \\
    + \frac{1}{2}\left[ {\left( {{x^2} + {y^2}} \right) + \nu \,{t^2}} \right]{\left| a \right|^2} - \frac{1}{2}{\left| a \right|^4},
\end{gather}
\noindent where the Cartesian coordinates are restored: $x = r\cos \left( \vartheta  \right),\,\,y = r\sin \left( \vartheta  \right)$ ($r = \sqrt {{x^2} + {y^2}}$ and $\vartheta$ are the radial and azimuthal cylindrical coordinates, respectively).

The driving ``force'' is defined by the dissipative terms in Eq. (\ref{eq:1}):

\begin{equation} \label{eq:4}
    Q =  i\left[ {-\Lambda \,a + \tau \,{\partial _{t,t}}a - \kappa \left( {{x^2} + {y^2}} \right)} \right]a.
\end{equation}

The Euler-Lagrange-Kantorovich equation reads as \cite{ANKIEWICZ200791}:

\begin{equation}
\label{eq:5}
\frac{{\delta \tilde L} }{{\delta {\rm{f}}}} - \frac{d}{{dZ}}\frac{{\delta \tilde L}}{{\delta {\rm{f}}}} = 2\Re\, \int\limits_{ - \infty }^\infty  {\int\limits_0^\infty  {\int\limits_0^{2\pi } {\,r\,Q\,\frac{{\delta a}}{{\delta {\rm{f}}}}\,dt\,dr\,d\vartheta}}}.
 \end{equation}

\noindent Here, the variation over the $Z-$dependent parameters ${\mathop{\rm f}\nolimits}  = \left( {\alpha ,\,\,\phi ,\,\,\theta ,\,\,\psi ,\,\,\upsilon ,\,\,\rho } \right)$ of (2) for the reduced Lagrangian

\begin{equation} \label{eq:6}
    \tilde L = \int\limits_{ - \infty }^\infty  {\int\limits_0^\infty  {\int\limits_0^{2\pi } {r\,L\,dt\,dr\,d\vartheta } } }
\end{equation}

\noindent was performed after substituting the ansatz (2) in $L$ and $Q$.

\section{\label{sec:results}Results}

\subsection{Spatiotemporal dissipative solitons}

From some algebra with (2,5) (see \cite{nb}), we obtained the following expressions for the 2D-DS parameters:

\begin{gather}
\label{eq:7}
    \theta = -\frac{\kappa  \rho^2}{2},\,\,\psi = \frac{120 \tau }{\pi ^2 \left(\sqrt{15} \sqrt{\upsilon^4 \left(15+128 \tau ^2\right)}+15\upsilon^2\right)},\nonumber\\
    \alpha^2= \frac{3 (1-\rho^4-\kappa ^2 \rho^8)}{\rho^2},\\
    \upsilon^2 = \frac{15 \left(\sqrt{1920 \tau ^2+225}-15\right)-64 \left(15+2 \pi ^2\right) \tau ^2}{384 \pi ^2 \tau  (\kappa  \rho^2+\Lambda )}\nonumber.
\end{gather}

The value of $\phi$ is irrelevant in our context. However, its value could be interesting for a stability analysis based on the Vakhitov-Kolokolov stability criterion \cite{malomed2016multidimensional}.

\noindent The remaining equation for the beam size $\rho$:

\begin{gather} \label{eq:8}
    \frac{\left(64 \left(15+2 \pi ^2\right) \tau ^2-15 A+225\right) \left(\kappa ^2 \rho^8+\rho^4-1\right)}{128 \pi ^2 \rho^2 \tau  (\kappa
   \rho^2+\Lambda )}\nonumber\\
   +\frac{\left(64 \left(15+2 \pi ^2\right) \tau ^2-15 A+225\right)^2 \nu }{147456 \pi ^2 \tau ^2 (\kappa  \rho^2+\Lambda )^2}-\\
   \frac{160
   \left(\left(3+\pi ^2\right) A -15\left(\pi^2-9\right)\right) \tau ^2}{\pi ^2 \left(A+15\right)^2}=2,\nonumber \\
   A= \sqrt{1920 \tau ^2+225}.\nonumber
\end{gather}

\noindent can be numerically solved.

\begin{figure}[htpb]
\includegraphics[width=8.8cm]{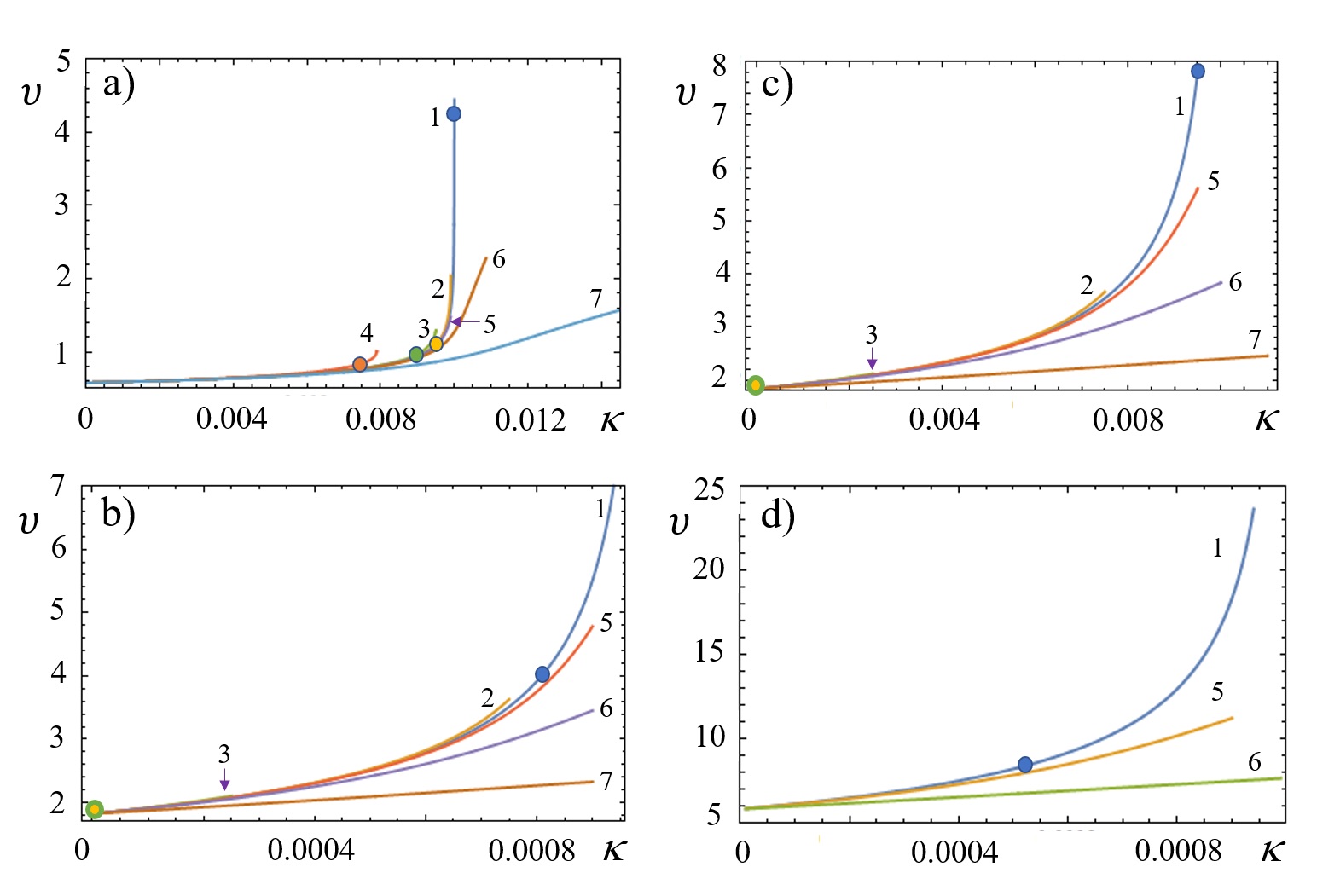}
\caption{\label{fig2} Dependencies of the DS temporal width $\upsilon$ on the dissipation gradient parameter $\kappa$. $\tau=$0.01 (a, b), 0.1 (c, d); $\Lambda=$-0.01 (a, c), -0.001 (b, d). The curves correspond to different values of the phase modulation parameter $\nu=$0 (1), 0.001 (2), 0.01 (3), 0.1(4), -0.001 (5), -0.01 (6), -0.1 (7). The colored points show the minimal $\kappa-$parameter, which is required for DS stabilization (colors refer to the color of the corresponding curve, see Table 1).}
\end{figure}

\begin{figure}[htpb]
\includegraphics[width=8.8cm]{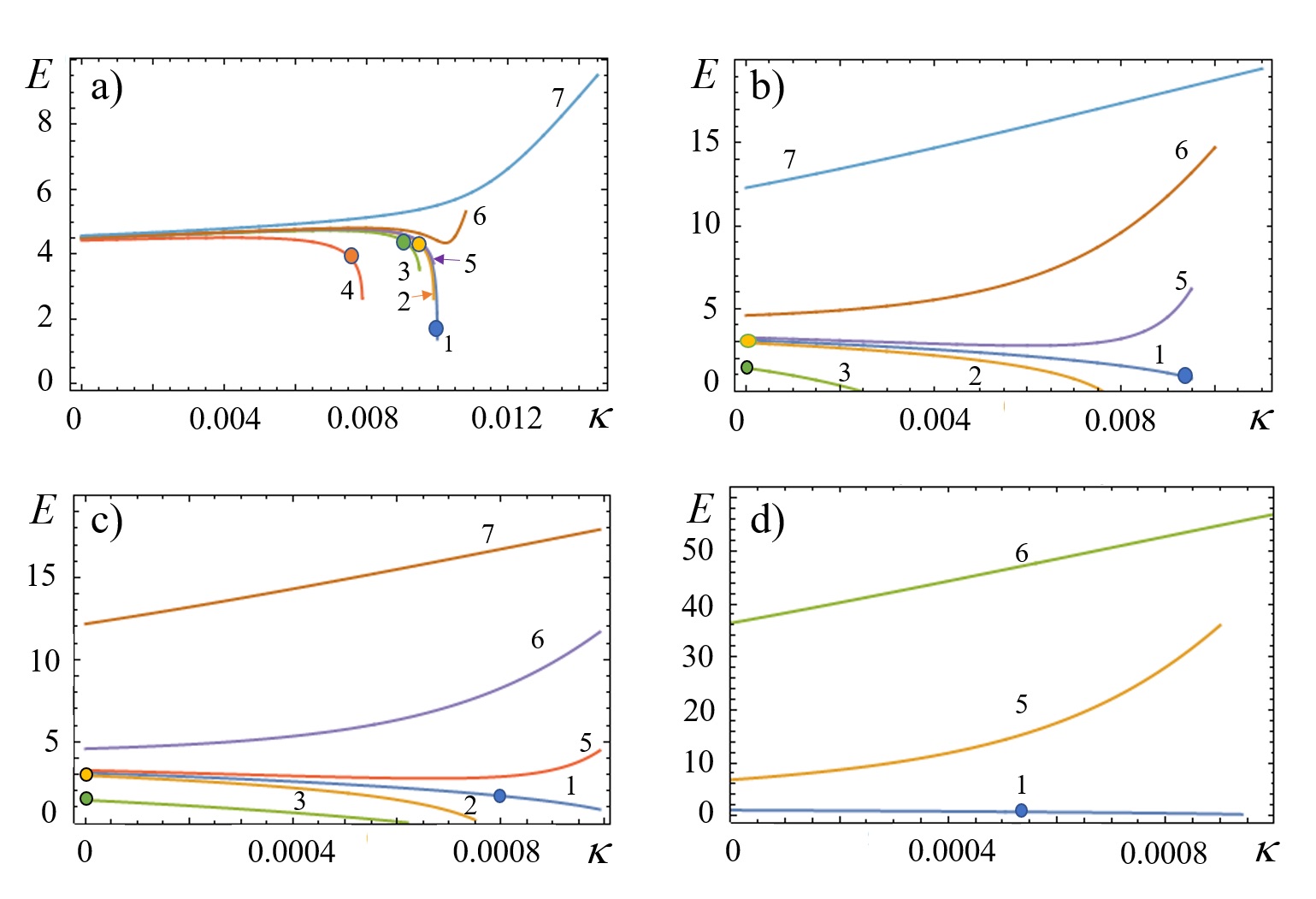}
\caption{\label{fig3} Dependencies of the DS energy $E$ on the dissipation gradient parameter $\kappa$. All remarks correspond to Fig. \ref{fig2}.}
\end{figure}

The DS width and energy dependencies on the dissipation gradient parameter $\kappa$, for different values of $\Lambda$, $\tau$, and $\nu$, are shown in Figs. \ref{fig2}, \ref{fig3}.

Figs. \ref{fig2}, \ref{fig3} demonstrate a clear division between the DS branches corresponding of a positive $\nu>0$ (``guiding'', curves 2 --4) or a negative $\nu<0$ (``anti-guiding'', curves 5 --7) phase modulation parameter, respectively. Such a difference is due to the self-focusing type of nonlinearity and the anomalous chromatic dispersion. Namely, $\nu>0$ contributes additively to the DS chirp, which reduces the overall nonlinear phase shift, in agreement with the chirp-free condition (see Appendix, Eq. (A2)):

\begin{equation}
\frac{2}{\upsilon^2}=\pi ^2 \nu \upsilon^2+\alpha^2
\end{equation}

\noindent As a consequence, $\nu>0$ leads to a decrease of the peak power and energy $E$ of the DS. One must note that the DS width remains almost unaffected in this case, but the range of DS existence vs. the $\kappa-$parameter shrinks.

The opposite situation takes place for a negative phase modulation coefficient $\nu<$0: here the DS energy $E$ increases, and simultaneously the DS width $\upsilon$ decreases.

Table \ref{tab:table1} demonstrates that, for a fixed value of the spectral dissipation coefficient $\tau$, a stable LB (see Section B) may be obtained within a broader range of the graded dissipation parameter $\kappa$, whenever the phase modulation coefficient $\nu>$0 grows larger. This means that both spectral diffusion and phase modulation are essential factors in LB stabilization. In principle, this effect could be connected with the known property of DS stabilization by spectral filtering that is obtained in chirped-pulse oscillators \cite{PhysRevA.79.043829}.

In our case, the ST confinement can be illustrated through the densities of energy generation in the radial ($P(r)$) and in the temporal ($P(t)$) dimensions, respectively. These are governed by the fluxes of energy,  $j(r)$ and $j(t)$, in the $r$ and $t$ directions for different $t$ and $r$ slices   \cite{akhmediev2008dissipative,maimistov1993evolution}:

\begin{equation}\label{eq:density}
    \begin{array}{l}
P(r) = \frac{{\partial j(r)}}{{\partial r}} = \frac{{2{\alpha ^2}\theta \,{e^{ - \frac{{{r^2}}}{{{\rho ^2}}}}}}}{{{\rho ^2}}}\left( {2{r^2} - {\rho ^2}} \right)\,\,{\mathop{\rm sech}\nolimits} \,{\left( {\frac{t}{\upsilon }} \right)^2},\\ \\
P(t) = \frac{{\partial j(t)}}{{\partial t}} = \frac{{2{\alpha ^2}\psi \,{e^{ - \frac{{{r^2}}}{{{\rho ^2}}}}}}}{\upsilon }\left( {2t\,\tanh \left( {\frac{t}{\upsilon }} \right) - \upsilon } \right).
\end{array}
\end{equation}

\begin{equation}\label{eq:flux}
\begin{split}
j(r) = \frac{i}{2}\left( {a\,\frac{{\partial {a^*}}}{{\partial r}} - {a^*}\,\frac{{\partial a}}{{\partial r}}} \right) =\\  - 2{\alpha ^2}{e^{ - \frac{{{r^2}}}{{{\rho ^2}}}}}r\,\theta \,{\mathop{\rm sech}\nolimits} \,{\left( {\frac{t}{\upsilon }} \right)^2},\\
j(t) = \frac{i}{2}\left( {a\,\frac{{\partial {a^*}}}{{\partial t}} - {a^*}\,\frac{{\partial a}}{{\partial t}}} \right) =\\  - 2{\alpha ^2}{e^{ - \frac{{{r^2}}}{{{\rho ^2}}}}}t\,\psi \,{\mathop{\rm sech}\nolimits} \,{\left( {\frac{t}{\upsilon }} \right)^2}.
\end{split}
\end{equation}

\noindent
One may see from Eqs. (\ref{eq:density}, \ref{eq:flux}) that a nontrivial energy redistribution inside the DS is produced by the presence of both temporal and spatial chirp.

\begin{figure}[htpb]
\includegraphics[width=8.8cm]{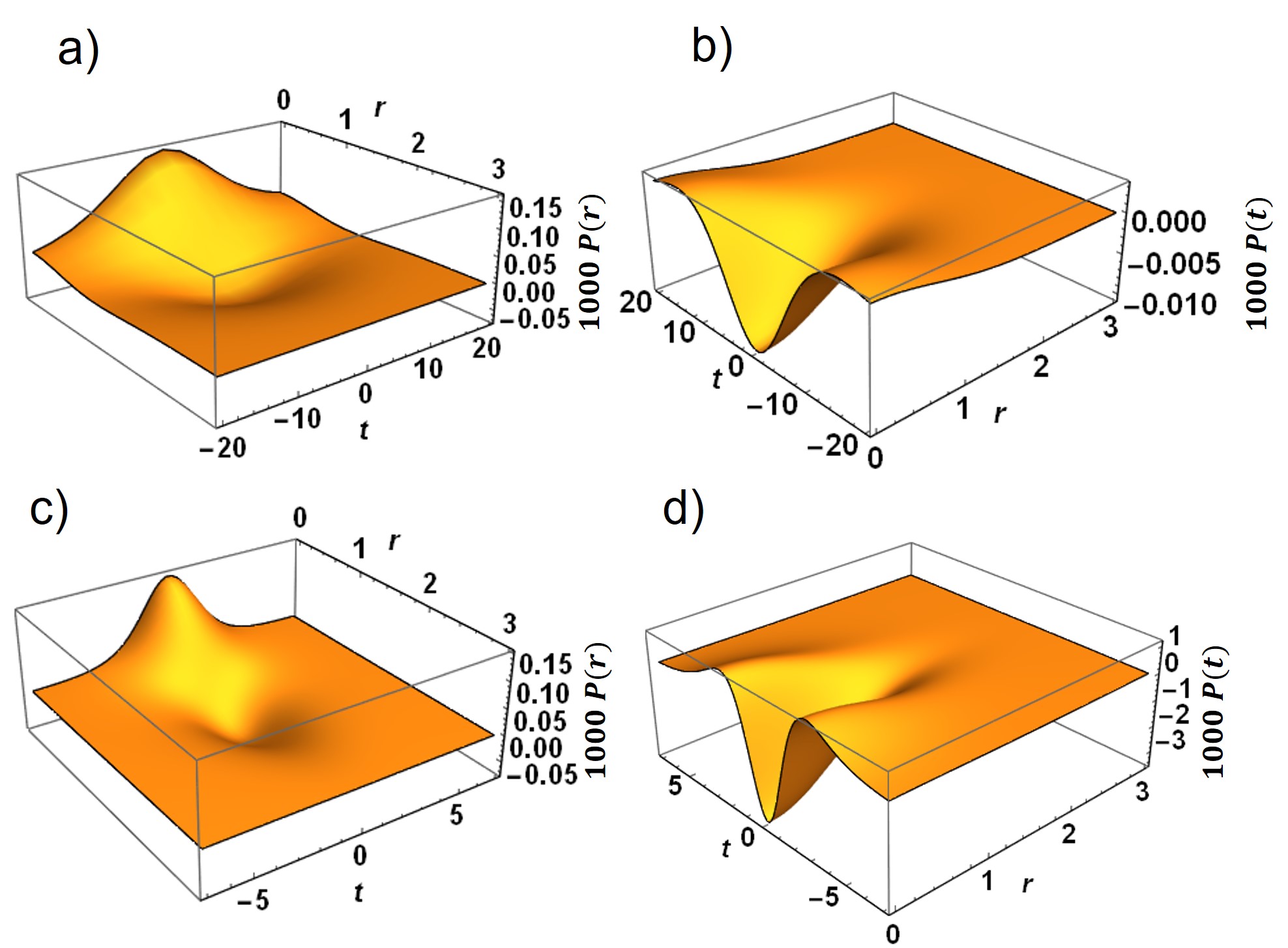}
\caption{\label{fig4} Densities of energy generation $P(r)$ and $P(t)$ in the $r-$ (a, c) and $t-$ (b, d) directions, respectively. $\kappa=$0.00975 (a, b) and 0.001 (c, d), $\nu=$0 (a, b) and 0.01 (c, d). $\Lambda=-$0.01, $\tau=$0.1.}
\end{figure}

\begin{figure}[htpb]
\includegraphics[width=8.8cm]{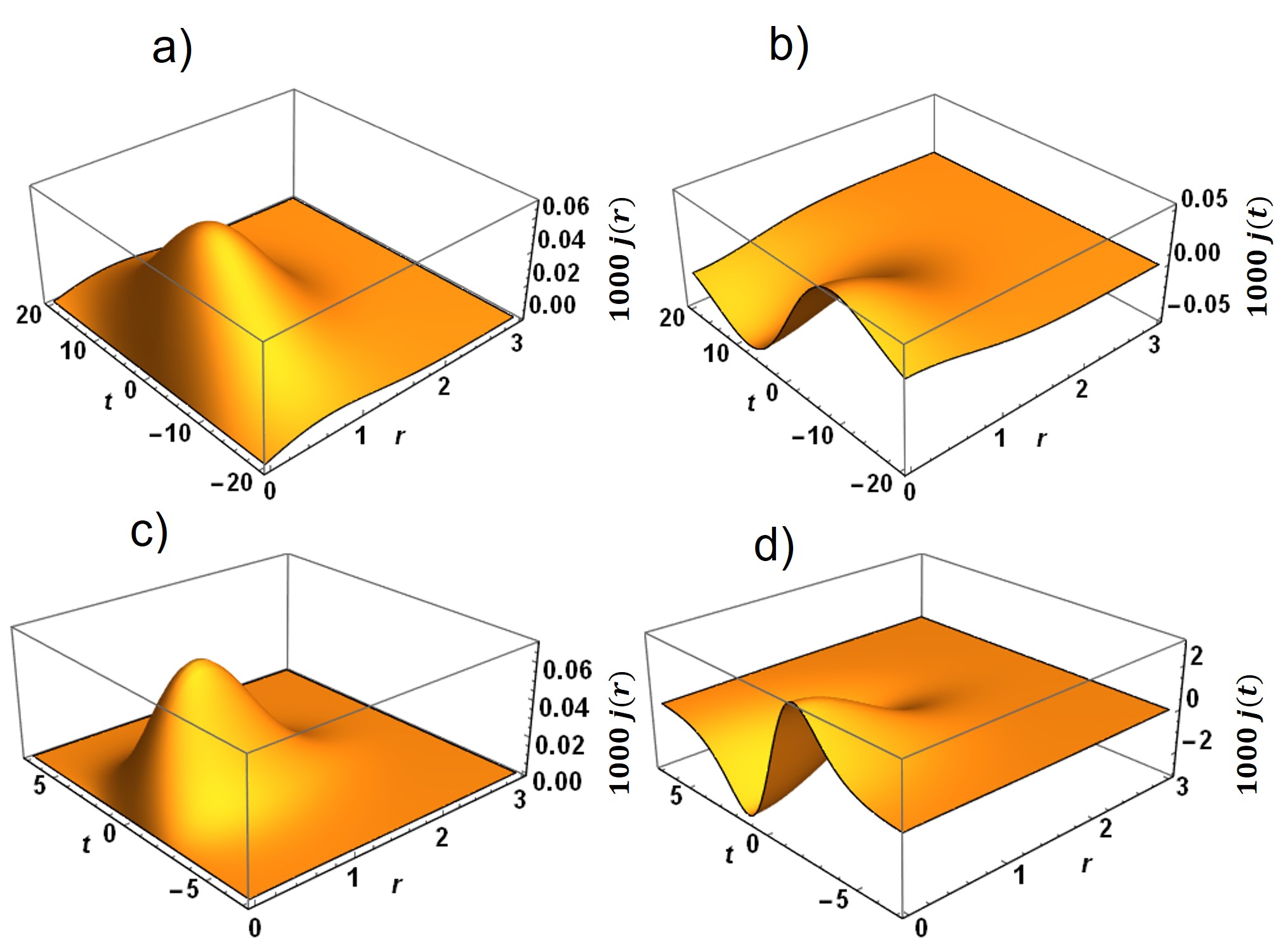}
\caption{\label{fig5} Energy fluxes $j(r)$ and $j(t)$ in the $r-$ (a, c) and $t-$ (b, d) directions, respectively. $\kappa=$0.00975 (a, b) and 0.001 (c, d), $\nu=$0 (a, b) and 0.01 (c, d). $\Lambda=-$0.01, $\tau=$0.1.}
\end{figure}

The densities of energy generation and their corresponding fluxes are shown in Figs. \ref{fig4}, \ref{fig5}, respectively, for either zero (a, b) or nonzero (c, d) phase modulation parameter $\nu$. The figures demonstrate confinement along the radial $r-$direction. This occurs whenever the energy which is generated on the $r=$0 axis (Figs. \ref{fig4} (a, c)) flows away from the center ($j(r)>$0, Figs. \ref{fig5} (a, c)) and is dissipated at some distance away from the beam axis, owing to the presence of graded dissipation. Simultaneously, the positive values of $j(t)$ for $t<$0, as well as the negative values of $j(t)$ for $t>$0 (see Figs. \ref{fig5} (b, d)) indicate the presence of an energy flow from the tails of the DS, where energy is generated, towards its center, where it is dissipated (Figs. \ref{fig4} (b, d)).

The impact of phase modulation $\nu  \ne$0 consists in the squeezing of the energy generation domain along the $t-$direction of $P(r)$ (see Fig. \ref{fig4} (c)), owing to the significant energy generation in the DS tails (see Fig. \ref{fig4} (d)), followed by the increased energy flux towards the middle of the DS (see Fig. \ref{fig5} (d)). Thus, one may conjecture that the enhancement of the energy generation and flux which is observed for $\nu>$0 will increase the DS robustness (Table \ref{tab:table1}).

\begin{figure}[htpb]
\includegraphics[width=8.8cm]{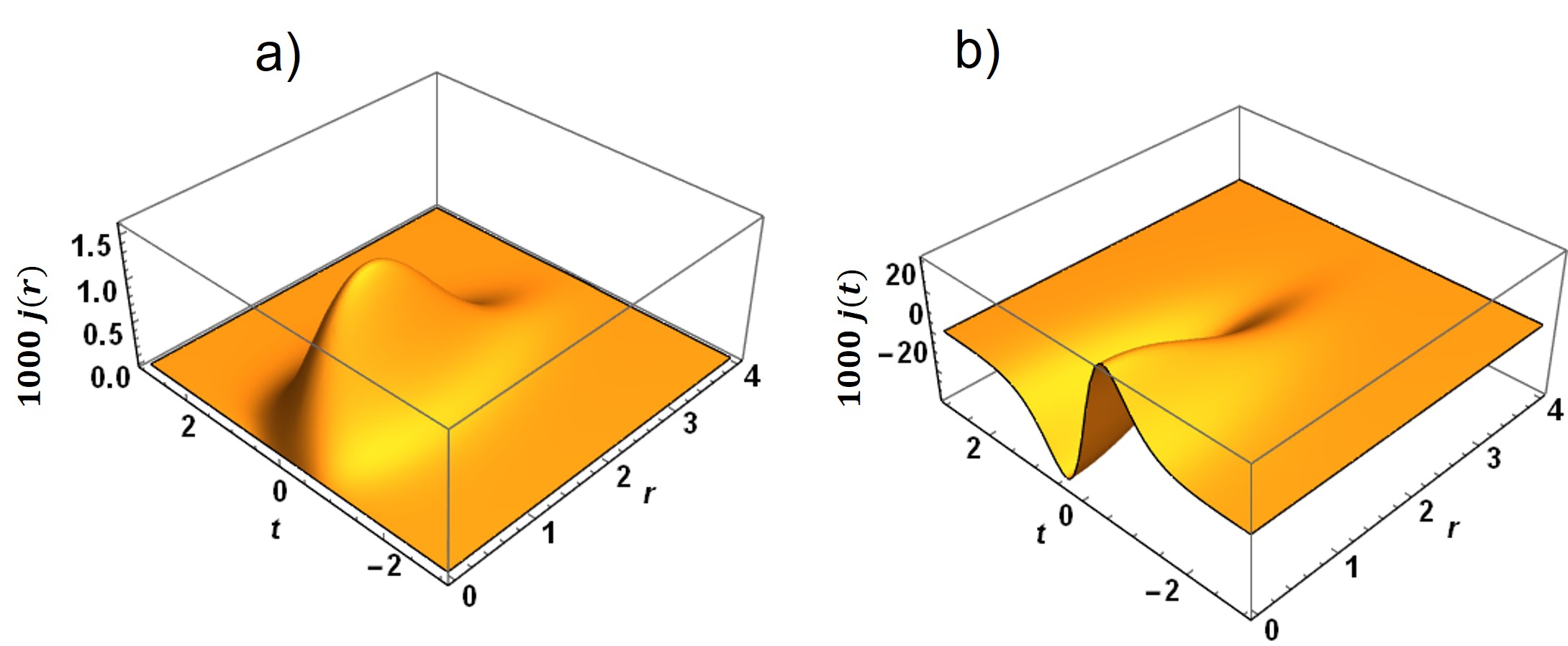}
\caption{\label{fig6} Energy fluxes $j(r)$ and $j(t)$ in the $r-$ (a) and $t-$ (b) directions, respectively. $\kappa=$0.001 , $\nu=$0.01, $\Lambda=-$0.01, and $\tau=$0.01.}
\end{figure}

Since the LB is affected by both spatial and temporal effects, the problem of avoiding the LB collapse arises \cite{kelley,dudley2014instabilities}. In the case considered in this work, self-focusing increases the effective graded gain, owing to beam concentration towards $r=$0. As a result, the occurrence of LB collapse might appear as the only unavoidable outcome. However, temporal compression of the DS leads to its spectral broadening. Then collapse can be blocked by the presence of spectral filtering (or ``kinetic cooling'' in BEC) \cite{our1,our2}. The comparison of Figs. \ref{fig5} (c, d) and Fig. \ref{fig6} demonstrates the occurrence of essential growth in energy generation and inward energy flux when the spectral dissipation is reduced (i.e., the $\tau-$parameter is decreased). Thus, one may assume that spectral dissipation can play a decisive role in DS stabilization, similarly to the case of high-energy femtosecond solid-state and fiber lasers \cite{Kalashnikov18}.

\begin{figure}[htpb]
\includegraphics[width=8.8cm]{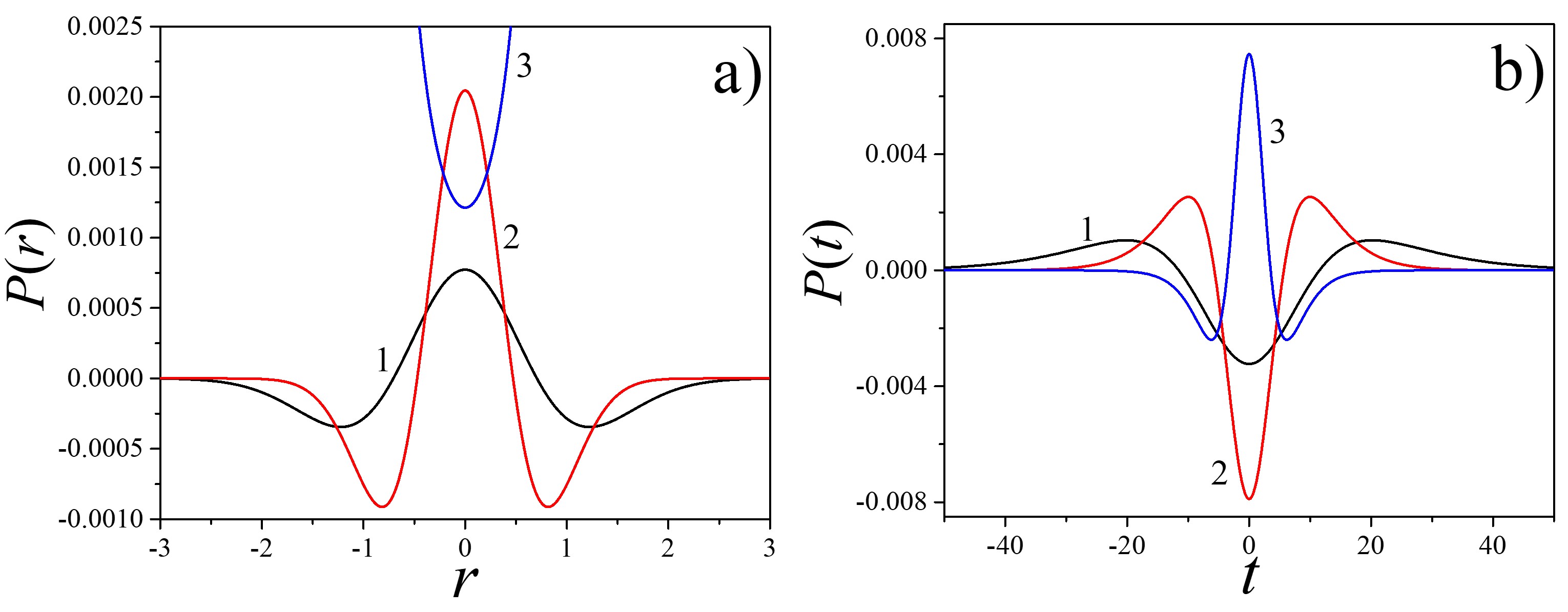}
\caption{\label{fig7} Densities of energy generation $P(r)_{t=0}$ and $P(t)_{r \to 0}$ in the $r-$ (a) and $t-$ (b) directions, respectively. $\nu=$0 (1), 0.01 (2), and -0.01 (3). Curves 1 are scaled $\times 100$ and $\times 1000$ in (a) and (b), respectively. $\Lambda=-$0.001, $\tau=$0.1, $\kappa=$0.00085.}
\end{figure}

One could suggest that the difference between the situations with either $\nu \ge 0$ or $\nu < 0$ follows from the different distributions of the energy generation densities (see Fig. \ref{fig7}). Specifically, 1) i$\nu < 0$: energy is generated at the center of the temporal profile of the DS and is dissipated at its tails (see curve 3 in Fig. \ref{fig7}, b); 2) i$\nu > 0$: energy is generated at the DS tails and is dissipated at the pulse center (see curves 1, 2 in Fig. \ref{fig7}, b). This dissimilarity corresponds to the separation between the so-called ``dissipative anti-solitons'' and ``dissipative solitons'', respectively \cite{ANKIEWICZ2007454}. These two branches crucially differ in the spatial domain: the anti-soliton eventually becomes unconfined (see curve 3 in Fig. \ref{fig7}, a), which entails the instability of such a structure. 

\subsection{Spatiotemporal soliton stability analysis}

\begin{figure}[htpb]
\includegraphics[width=8.8cm]{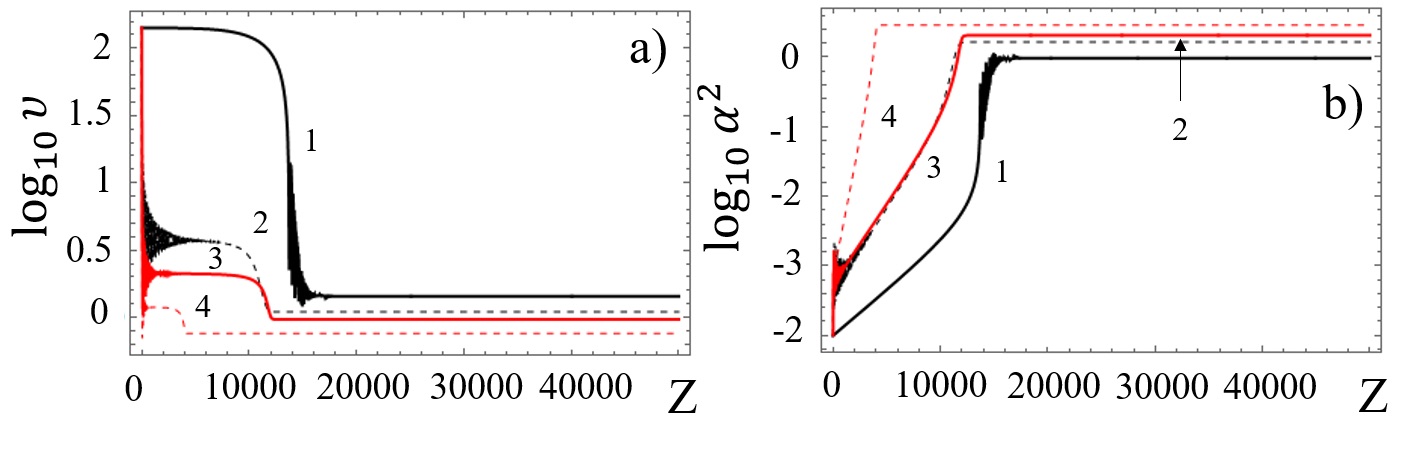}
\caption{\label{fig8} Evolution of the DS temporal width (a) and intensity (b), in the vicinity of the stability threshold (i.e., the minimal $\kappa$, as pointed by circles in Figs. \ref{fig2}, \ref{fig3}), for the parameters of Figs. \ref{fig2}, \ref{fig3}, a.}
\end{figure}

\begin{figure}[htpb]
\includegraphics[width=8.8cm]{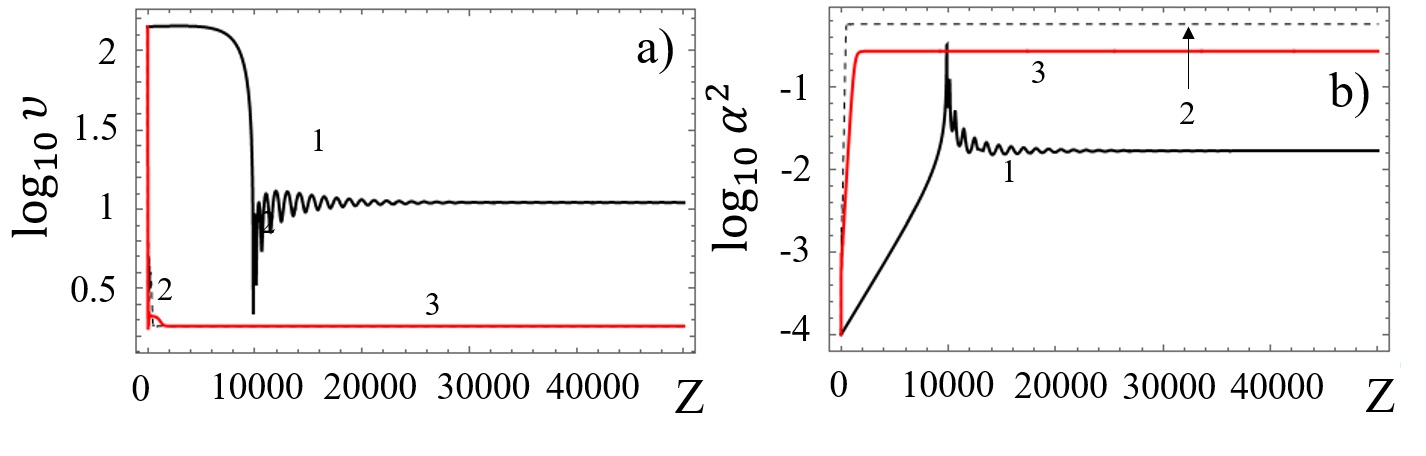}
\caption{\label{fig9} Evolution of the DS temporal width (a) and intensity (b), in the vicinity of stability threshold (i.e., minimal $\kappa$, as pointed by circles in Fig. \ref{fig2}, \ref{fig3}), for the parameters of Fig. \ref{fig2}, \ref{fig3}, c.}
\end{figure}

\begin{figure}[htpb]
\includegraphics[width=8.8cm]{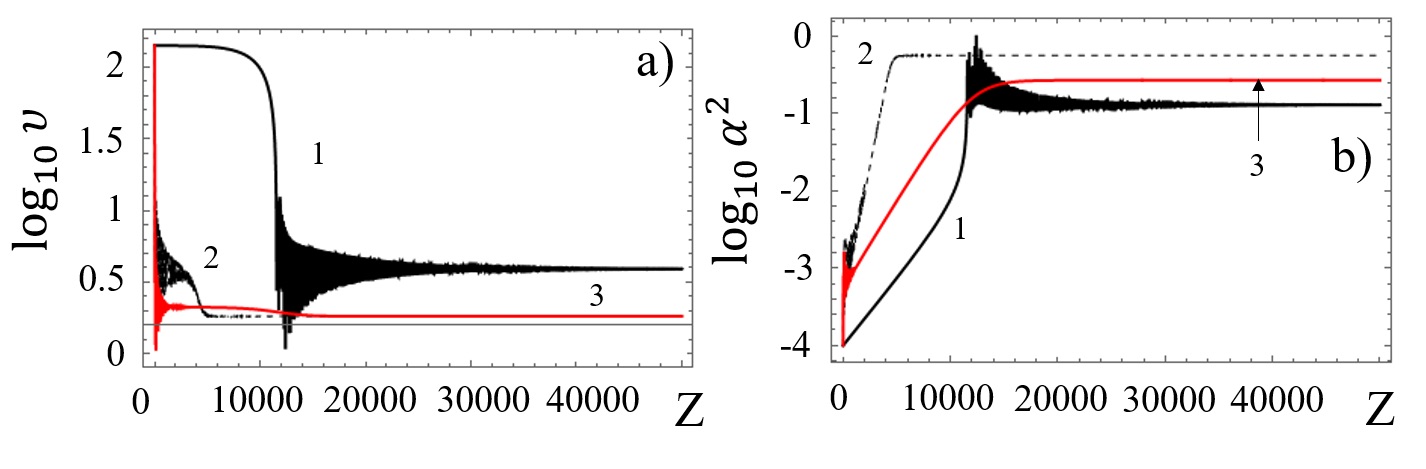}
\caption{\label{fig10} Evolution of the DS temporal width (a) and intensity (b), in the vicinity of stability threshold (i.e., minimal $\kappa$, as pointed by circles in Fig. \ref{fig2}, \ref{fig3}), for the parameters of Fig. \ref{fig2}, \ref{fig3}, b.}
\end{figure}

\begin{figure}[htpb]
\includegraphics[width=8.8cm]{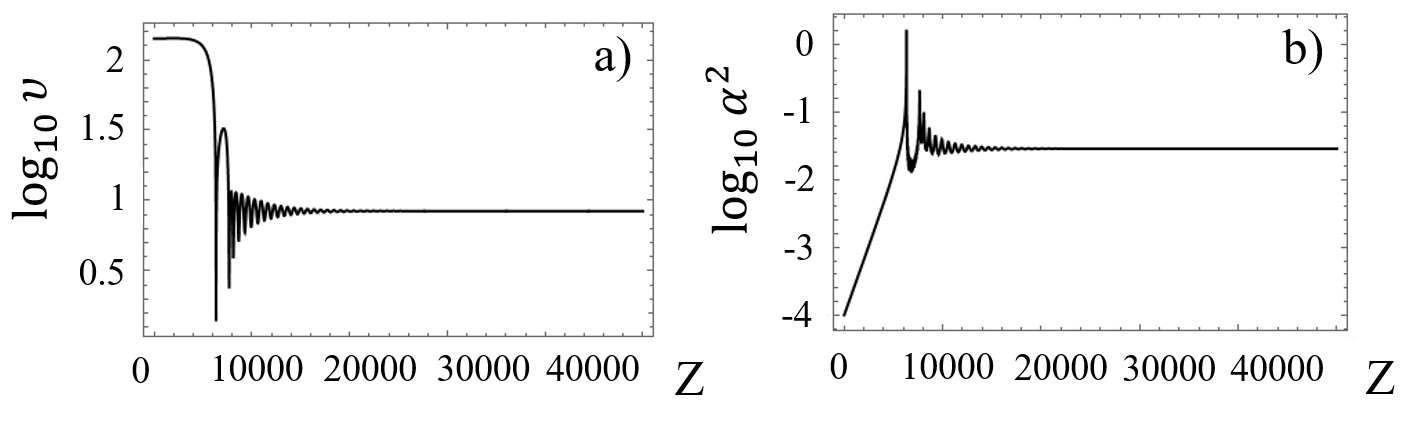}
\caption{\label{fig11} Evolution of the DS temporal width (a) and intensity (b) in the vicinity of the stability threshold, for $\tau=$0.1, $\Lambda=-$0.001, $\nu=$0.01, and $\kappa=$0.000525.}
\end{figure}

To analyze the stability of solutions (7,8), we numerically evaluated the generating system (A1 -- A5) (see Appendix). The calculated evolution of the DS width and intensity in the vicinity of the stability borders, marked by circles in Figs. \ref{fig2}, \ref{fig3}, are illustrated by Figs. \ref{fig8}--\ref{fig11}. Table \ref{tab:table1} summarizes the stability properties obtained from these simulations. The initial conditions on $Z=$0 are $\alpha_0 = 0.01,\,\,\theta_0 = \psi_0 = 0,$ $\upsilon_0 = {{\sqrt 2 } \mathord{\left/
 {\vphantom {{\sqrt 2 } {\alpha_0}}} \right.
 \kern-\nulldelimiterspace} {\alpha_0}}$, ${\rho _0} = {{\left( {{\alpha _0} - \sqrt {\alpha _0^4 + 36} } \right)} \mathord{\left/
 {\vphantom {{\left( {{\alpha _0} - \sqrt {\alpha _0^4 + 36} } \right)} 6}} \right.
 \kern-\nulldelimiterspace} 6}$ \cite{our1}.

\begin{table*}[htpb]
\caption{\label{tab:table1} Stability properties of DSs, and the corresponding minimum graded dissipation parameter $\kappa_{min}$.}
\begin{ruledtabular}
\begin{tabular}{cccc}
  $\Lambda$ & $\tau$\ & $\nu$ & $\kappa_{min}$ \\
    \hline
  -0.01 & 0.01 & 0 & 0.00985 \\
  -0.01 & 0.01 & 0.001 & 0.0095 \\
  -0.01 & 0.01 & 0.01 & 0.009 \\
  -0.01 & 0.01 & 0.1 & 0.00675 \\
  -0.01 & 0.1 & 0 & 0.00975 \\
  -0.01 & 0.1 & 0.001 & 0 \\
  -0.01 & 0.1 & 0.01 & 0 \\
  -0.01 & 0.1 & 0.1 & no solutions \\
  -0.001 & 0.01 & 0 & 0.0008 \\
  -0.001 & 0.01 & 0.001 & 0 \\
  -0.001 & 0.01 & 0.01 & 0 \\
  -0.001 & 0.01 & 0.1 & no solutions \\
  -0.001 & 0.1 & 0 & 0.000525 \\
  -0.001 & 0.1 & 0.001 & no stable solutions \\
  -0.001 & 0.1 & 0.01 & no stable solutions \\
  -0.001 & 0.1 & 0.1 & no solutions \\
  -0.01 & 0.01 & $\nu<0$ & no stable solutions \\
  -0.01 & 0.1 & $\nu<0$ & no stable solutions \\
  -0.001 & 0.01 & $\nu<0$ & no stable solutions \\
  -0.001 & 0.1 & $\nu<0$ & no stable solutions \\
\end{tabular}
\end{ruledtabular}
\end{table*}

From our numerical analysis, we may note the following three main results: i) there is a maximum phase modulation parameter $\nu$, for which there is no DS solution, ii) there is no DS for $\nu<$0, iii) the action of phase modulation ($\nu \ne $0) broadens the DS stability range with respect to the graded dissipation parameter $\kappa$, down to its zero level (i.e., a gain/loss profiling is no longer required in this case).

Figs. \ref{fig8}--\ref{fig11} demonstrate the numerically computed evolution of the DS temporal width $\upsilon$ and intensity $\alpha^2$, as a function of propagation distance $Z$. The calculations show the presence of three types of instabilities: i) a breathing behavior in the vicinity of the threshold $\kappa \buildrel\textstyle.\over= \kappa_{min}$, ii) LB collapse, and iii) LB decay. However, the fundamental mode approximation cannot intrinsically describe the possible excitation of higher-order spatial modes. Therefore, direct numerical solutions of Eq. (1) are required. 

As a matter of fact, Figs. \ref{fig8}--\ref{fig11} illustrate the fact that the evolution of the DS parameters from arbitrary initial conditions deviates substantially from the adiabatic stationary solutions. Under these conditions, the convergence to a different stationary solution can still be considered as a manifestation of its global stability, akin to the capability of mode-locking self-starting. However, such convergence can be highly nonadiabatic (e.g., see the transient peak intensity bursts in Figs. \ref{fig9} (b) and \ref{fig11} (b)). This effect, i.e., the transient Q-switching-like intensity fluctuations, was experimentally observed in a solid-state laser with distributed Kerr-lens mode-locking \cite{brons2014energy}. These results indicate that the transient shape of an evolving LB can differ essentially from the adiabatic ansatz. Therefore, any definitive conclusion about the LB stability requires carrying out full numerical simulations in the framework of either 2D and 3D versions of Eq. (\ref{eq:1}). In the following sections, we present the results of such simulations.

\subsection{Numerical soliton solutions}

In our numerical simulations of Eq. (1), we used the COMSOL 5.4 software and the generalized alpha finite-element method (FEM) as implemented by the PARDISO solver on a free quad mesh with approximately 6300 quads. The step size and the propagation distance were $\Delta \xi  = 0.1$ and $L=10000$, respectively, when using the normalization of Eq. (\ref{eq:1}). A typical simulation window was $\left[ {t, - 100,100} \right] \times [r,{10^{ - 6}},10]$. A free tetrahedral mesh with approximately 98000 tetrahedra was used for 3D simulations.

\subsection{Soliton stability without phase modulation}

\begin{figure}[htpb]
\includegraphics[width=9cm]{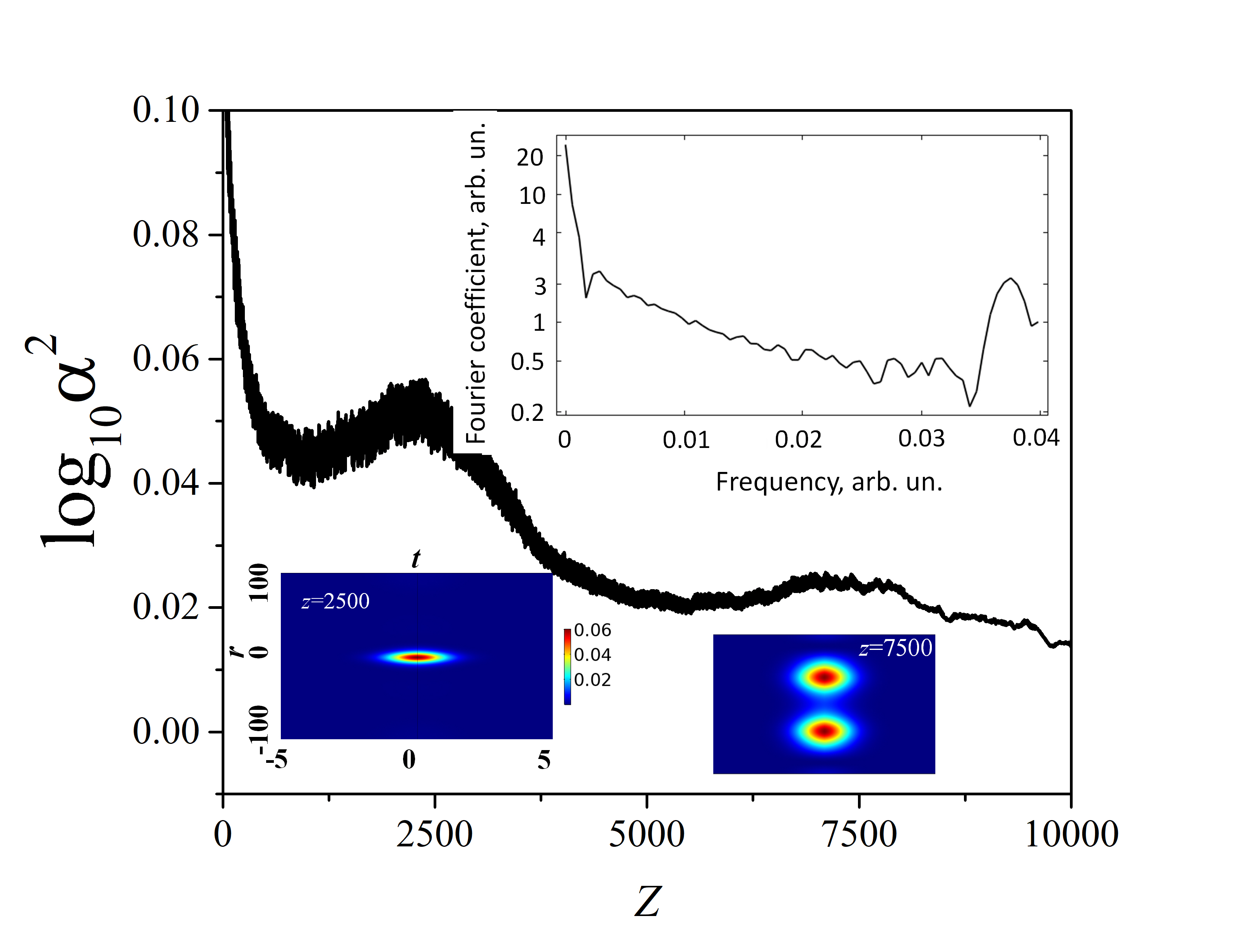}
\caption{\label{fig12} Averaged evolution of the maximum power at $r \to 0$ for $\tau=$0.1, $\Lambda=-$0.001, $\nu=$0, and $\kappa=$0.0007; $const=0.3$. Upper inset shows the spectrum of maximum intensity oscillations (in logarithmic scale) within the interval $Z \in \left[ {0,10000} \right]$. Lover insets show the contour plots of the LB power at $Z=$2500 and 7500.}
\end{figure}

\begin{figure}[htpb]

        \centering
     \includegraphics[width=8.75cm]{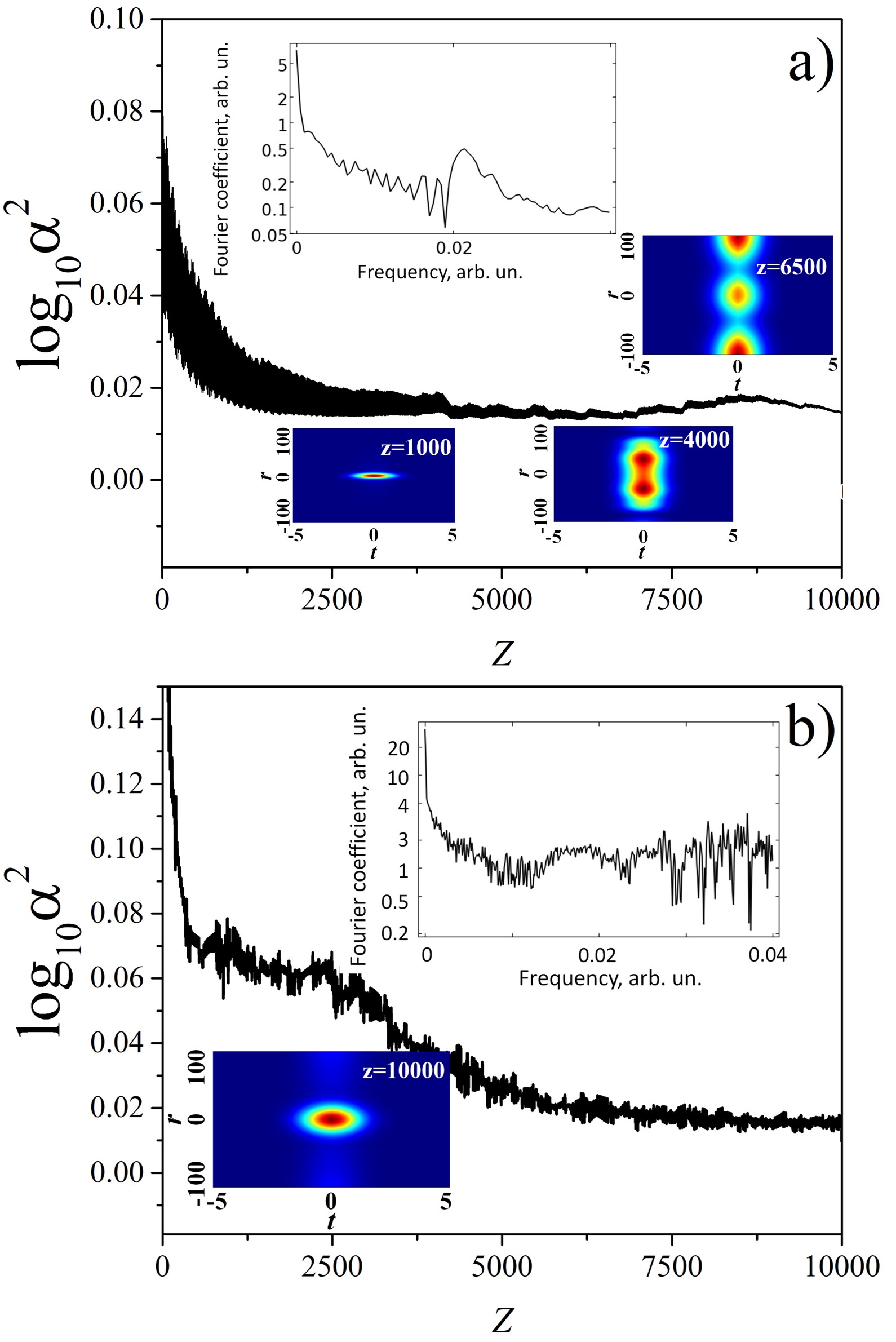}

        \caption{\label{fig13} Averaged evolution of the maximum power at $r \to 0$  for the different initial conditions (see Eq. (B1), Appendix)) of $const=0.3$ (a, the animated contour plot is available on \url{http://info.tuwien.ac.at/kalashnikov/fig13a.mp4}) and $1$ (b,  the animated contour plot is available on \url{http://info.tuwien.ac.at/kalashnikov/fig13b.mp4}). $\tau=$0.1, $\Lambda=-$0.001, $\nu=$0, and $\kappa=$0.00085. Insets show the contour plots of LB power at the indicated $Z$, and the spectra of maximum intensity oscillations (in logarithmic scale) within the interval $Z \in \left[ {0,10000} \right]$.}
        \end{figure}

First of all, numerical calculations confirm the absence of stable LB for $\nu < 0$. The amplitude of any initial pulse decays exponentially with distance $z$. Next, our analysis demonstrates that the spatially composed structure of an ST DS manifests itself in the form of ST breathing of the LB when $\kappa$ tends to the stability threshold, which is of $\kappa \approx 0.0005$ and calculated from (A1--A5) (see Table 1 and Appendix) for the parameters of Fig. \ref{fig12}. The LB dynamics exhibits spatio-temporal oscillations, causing a long modulated tail in the spectrum of the maximum intensity, as it is calculated over the propagation interval $Z=10000$ (see the upper inset in Fig. \ref{fig12}). The key deviation from the analytical results is the emission of a periodically oscillating ``radiation'' along the $t-$dimension (as shown in the bottom inset of Fig. \ref{fig13}, b), which eventually leads to LB splitting, i.e., to a multi-pulsing instability (see right bottom inset in Fig. \ref{fig12}).

Fig. \ref{fig12} demonstrates the impact of this instability, which is not grasped by the analytical model, namely, LB splitting in the time domain. Nevertheless, an insight in this temporal LB splitting can be obtained by exploiting the analytical results for (1+1)-dimensions (i.e., when limiting the analysis to the $Z-t$- coordinates) \cite{Kalashnikov18}. In this case, it is possible construct a DS stability region, which is provided by the so-called ``master diagram'', defined by the control parameter $C = {\tau  \mathord{\left/
 {\vphantom {\tau  \varsigma }} \right.
 \kern-\nulldelimiterspace} \varsigma }$ in our normalization (see \cite{Kalashnikov18} and the references therein; the parameter $\varsigma$ will be defined below). In the present case, the nonlinear mechanism providing LB formation is based on a combination of beam self-focusing with spatially graded dissipation (see Eq. (\ref{eq:1})). However, when reducing the number of dimensions, i.e., excluding the transverse ($r$) coordinate, it is necessary for DS existence to add an effective (or ``emergent'') dissipative nonlinear term (``self-amplitude modulation'', SAM) to the $r-$independent $Q$ in Eq. (\ref{eq:4}), say \cite{nb2,saha2019variational}  
 \begin{equation}\label{SAM}
  SAM = i{{\mu \varsigma {{\left| a \right|}^2}a} \mathord{\left/
 {\vphantom {{\mu \varsigma {{\left| a \right|}^2}a} {\left( {1 + \varsigma {{\left| a \right|}^2}} \right)}}} \right.
 \kern-\nulldelimiterspace} {\left( {1 + \varsigma {{\left| a \right|}^2}} \right)}}\approx i\mu \varsigma \left( {{{\left| a \right|}^2} - \varsigma {{\left| a \right|}^4}} \right)a.    
 \end{equation}
 \noindent Here $\mu$ is a modulation depth which is included in $\Lambda$, and $\varsigma$ is an inverse saturation power of the SAM.
 
Although the ansatz (2) does not apply to the description of multiple pulse generation, using a low-dimensional model permits us to obtain the stability criterion for the DS: there is a maximal value of $C$, which divides the regions of either single or multiple DSs formation. This maximal $C$ value is defined by the equation \cite{Kalashnikov18,nb2}:
 
 \begin{equation}\label{eq:threshold}
     \frac{2}{3}{\alpha ^2}C = \mu \left( {2 + \frac{{\ln \left( {\frac{{1 + {\alpha ^2} - \alpha \sqrt {1 + {\alpha ^2}} }}{{1 + {\alpha ^2} + \alpha \sqrt {1 + {\alpha ^2}} }}} \right)}}{{\alpha \sqrt {1 + {\alpha ^2}} }}} \right),  
 \end{equation}
\noindent which gives the following asymptotic expressions for the maximum $C$ confining the stability region: $\lim_{E \to 0} C = 0.1$, and $\lim_{E \to \infty} C_{\Sigma \to 0} \approx 3.5{{\sqrt {\mu \,\tau } } \mathord{\left/
 {\vphantom {{\sqrt {\mu \,\tau } } {\varsigma E}}} \right.
 \kern-\nulldelimiterspace} {\varsigma E}}$, respectively. In our case, these expressions only have a heuristic character. Specifically, we may derive the approximate dimensional inequality 
 \begin{equation}\label{eq:stab}
     \frac{\beta }{{\gamma E}}\sqrt {\frac{\mu }{\tau }}  \ge const
 \end{equation}
for describing the asymptotic stability threshold ($\gamma$ is the SPM coefficient, $\beta$ is the group-delay dispersion coefficient). 	

The above expressions for the stability threshold defined by both the $C-$ parameter and the DS energy $E$ are oversimplified but provide some helpful insight into the mechanics of the stability enhancement caused by the increase of the phase modulation depth and/or the $\tau-$decrease for a fixed energy value. However, in order to use this simple analogy with the (1+1)-dimensional case, the $\mu-$, and $\varsigma-$parameters must be connected with the transverse (i.e., $r-$dependent) parameters of Eqs. (\ref{eq:1}), (2). The simplest model for such a connection is presented in the Appendix.

From \cite{our1} and Appendix (C5), one may conjecture that $\mu  \propto {{{D^{ - 2}} = \kappa } \mathord{\left/
 {\vphantom {{{D^{ - 2}} = \kappa } {\left| \Lambda  \right|}}} \right.
 \kern-\nulldelimiterspace} {\left| \Lambda  \right|}}$. Hence, the $\kappa-$decrease and the $E-$ growth (Fig. \ref{fig3}) could violate the criterion (\ref{eq:stab}).
 
The numerical results illustrating the effects of increasing $\kappa$ (i.e., the dissipation grading enhancement) are shown in Fig. (\ref{fig13}). The transition to a steady-state becomes progressively smoother compared to the case of Fig. (\ref{fig12}). The peak power breathing, accompanied by a long high-frequency tail in the spectrum, is caused by the evanescent excitation of higher-order spatial modes. Also, Figs. \ref{fig13} ($a$, $b$) demonstrate a dependence of the dynamics on the initial value of $\alpha_0$ (see Appendix, (B1)). These results reveal the presence of two scenarios for the deviation from solution (2), which lead to its destabilization. Both scenarios are connected with a temporal deconfinement mechanism: either LB splitting (see Fig. \ref{fig12}, a) or generation of radiation tails in the time-domain (see Fig. \ref{fig13}, b).

\begin{figure}[htpb]
\includegraphics[width=9.6cm]{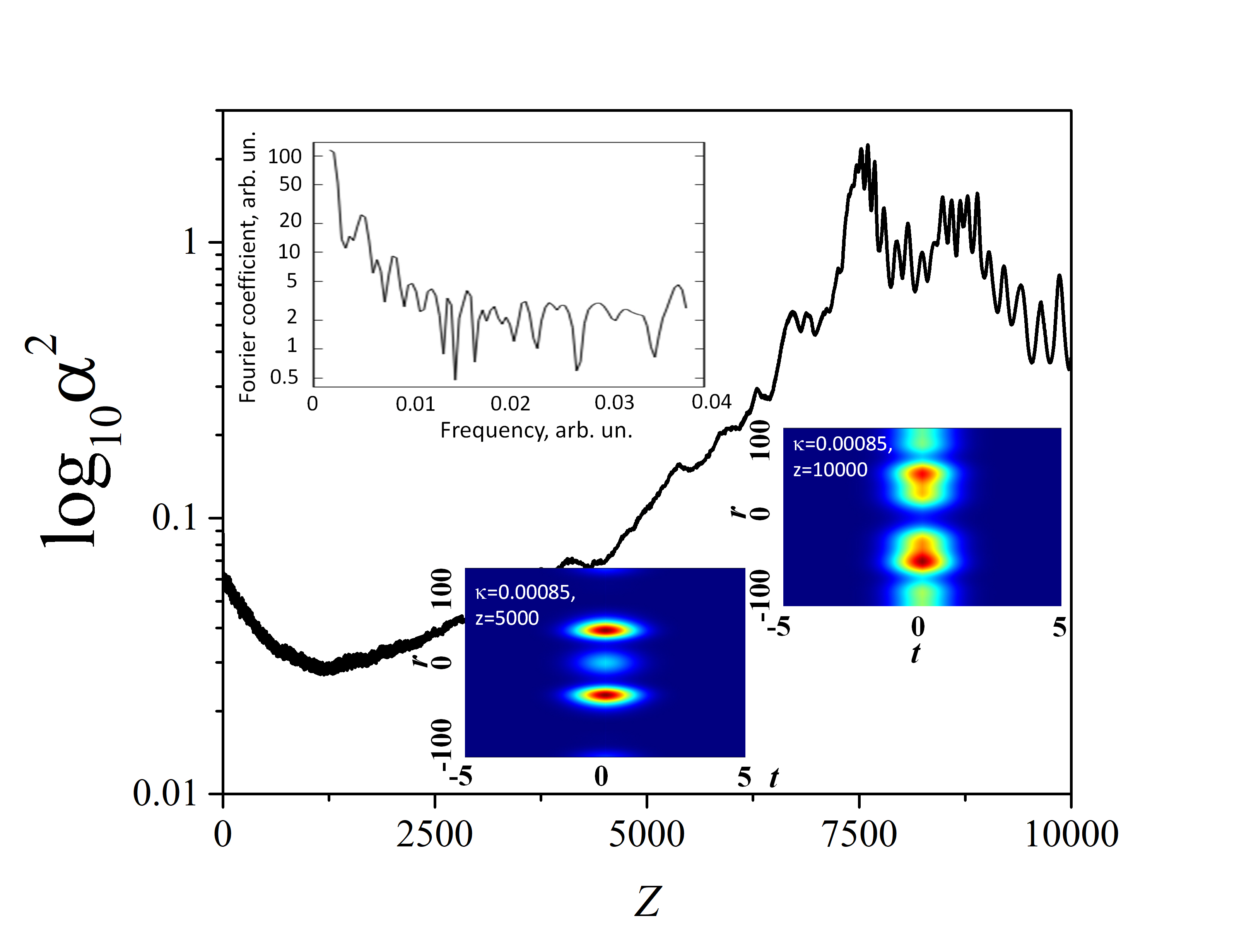}
\caption{\label{fig14} Averaged evolution of the maximum power at $r \to 0$ for $\tau=$0.01, $\Lambda=-$0.001, $\nu=$0, and $\kappa=$0.00085. The bottom insets show the contour plots of LB at $Z=$5000 (left) and $Z=$10000 (right). The upper inset shows the spectrum of maximum power oscillations (in logarithmic scale) within the interval $Z \in \left[ {0,10000} \right]$. $const=0.3$ in Eq. (B1) (see Appendix).}
\end{figure}

As it was pointed out before \cite{our1}, spectral dissipation (``kinetic cooling'') is a necessary condition for the existence of ST DSs. Table 1 demonstrates the soliton stability range squeezing when the spectral dissipation ($\tau-$parameter in Eq. (1)) decreases. Numerical simulations confirm this claim. For instance, Fig. \ref{fig14} shows the strong irregular dynamics of the maximum peak power and the progressive LB fragmentation and shape distortion which occurs when spectral dissipation decreases.

\begin{figure}[htpb]
\includegraphics[width=9.6cm]{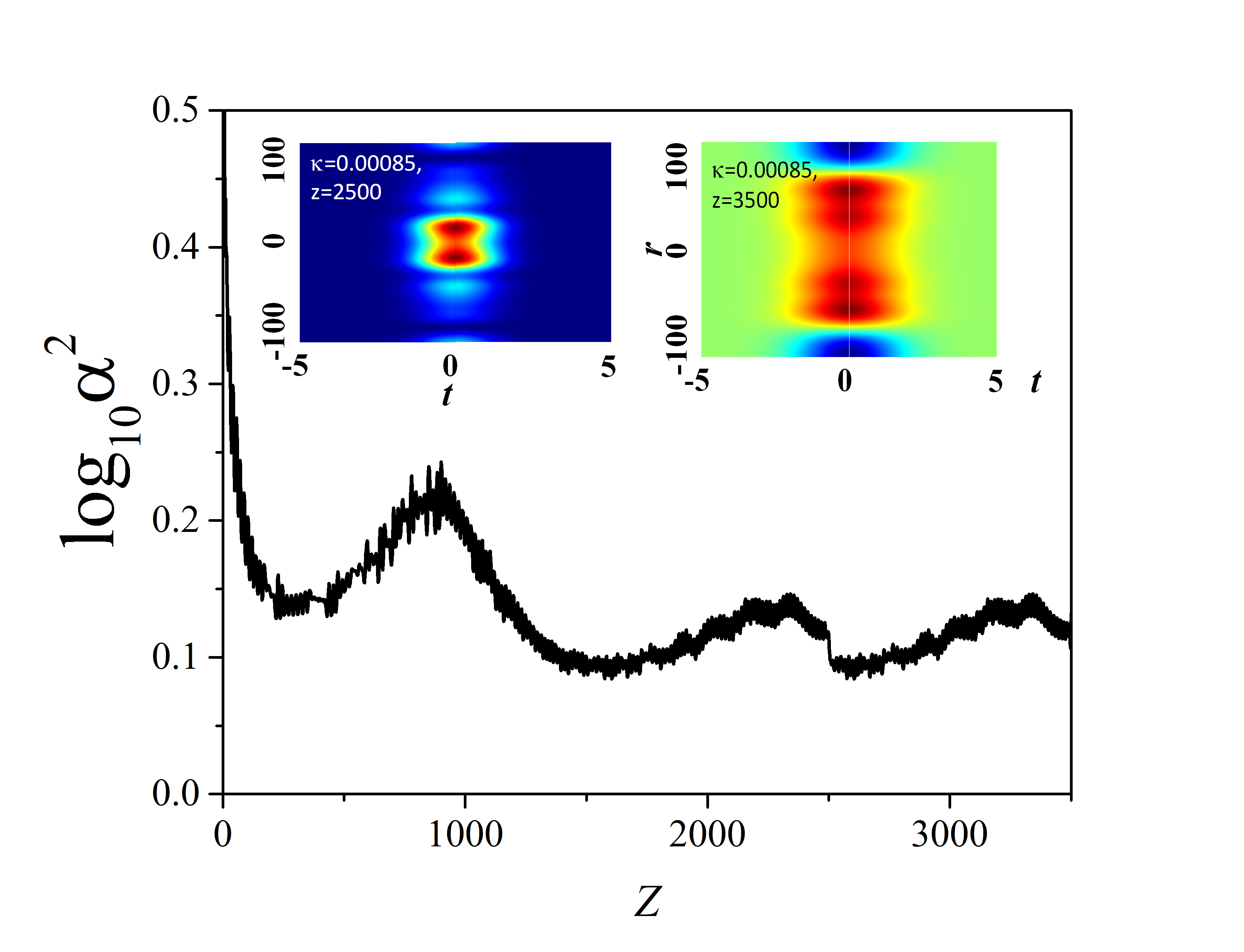}
\caption{\label{fig15} Averaged evolution of the maximum power at $r \to 0$ for $\tau=$0.01, $\Lambda=-$0.001, $\nu=$0, and $\kappa=$0.00085. The insets show the contour plots of LB at $Z=$2500 (left) and $Z=$3500 (right). $const=1$ in Eq. (B1) (see Appendix).}
\end{figure}

An additional aspect of ST DS destabilization induced by the decrease of spectral dissipation is connected with its dependence on initial conditions. Fig. \ref{fig15} demonstrates that, as the value of $\tau$ is reduced, temporal splitting of the DS becomes stronger whenever $\alpha_0$ grows larger until a loss of confinement occurs in temporal and spatial dimensions.

The results of previous numerical simulation suggest that pure spatial confinement does not guarantee perfect LB integrity in the $t-$domain. Moreover, the mixing of the ST degrees of freedom does not permit inferring a conclusive prediction about the LB stability when starting from the effective 1D-model (which, e.g., is misleading in predicting a decrease of the stability threshold $\propto 1/\sqrt{\tau}$ in Eq. (\ref{fig14})).

In order to better assess our ST-confinement scheme, given the complexity of the underlying instabilities, it is necessary to approach the ST confinement utilizing full numerical simulations, which is the subject of the next section.

\subsection{Soliton stability with phase modulation}

The presence of ``external phase-modulation'', whose depth is measured by the parameter $\nu$ in Eq. (\ref{eq:1}), introduces a temporal (i.e., $t-$dependent) localization of the phase ($\propto t^2+H.O.T.$), which corresponds to a 3D or pancake-like confining potential (see Fig. \ref{fig1}, $b$). In a fiber setup, such modulation can be realized, for instance, by a periodical phase modulation along a fiber, which is equivalent to the so-called phase active mode-locking \cite{1076343,Chang09} or vector resonance mode locking \cite{Sergeyev}.

Intuitively, $\nu<0$ breaks ST confinement, so that there is no stable ST DS (see Table \ref{tab:table1}). Numerical simulations confirm this statement: the LB-like initial seed decays exponentially in this case.

\begin{figure}[htpb]
\includegraphics[width=9.5cm]{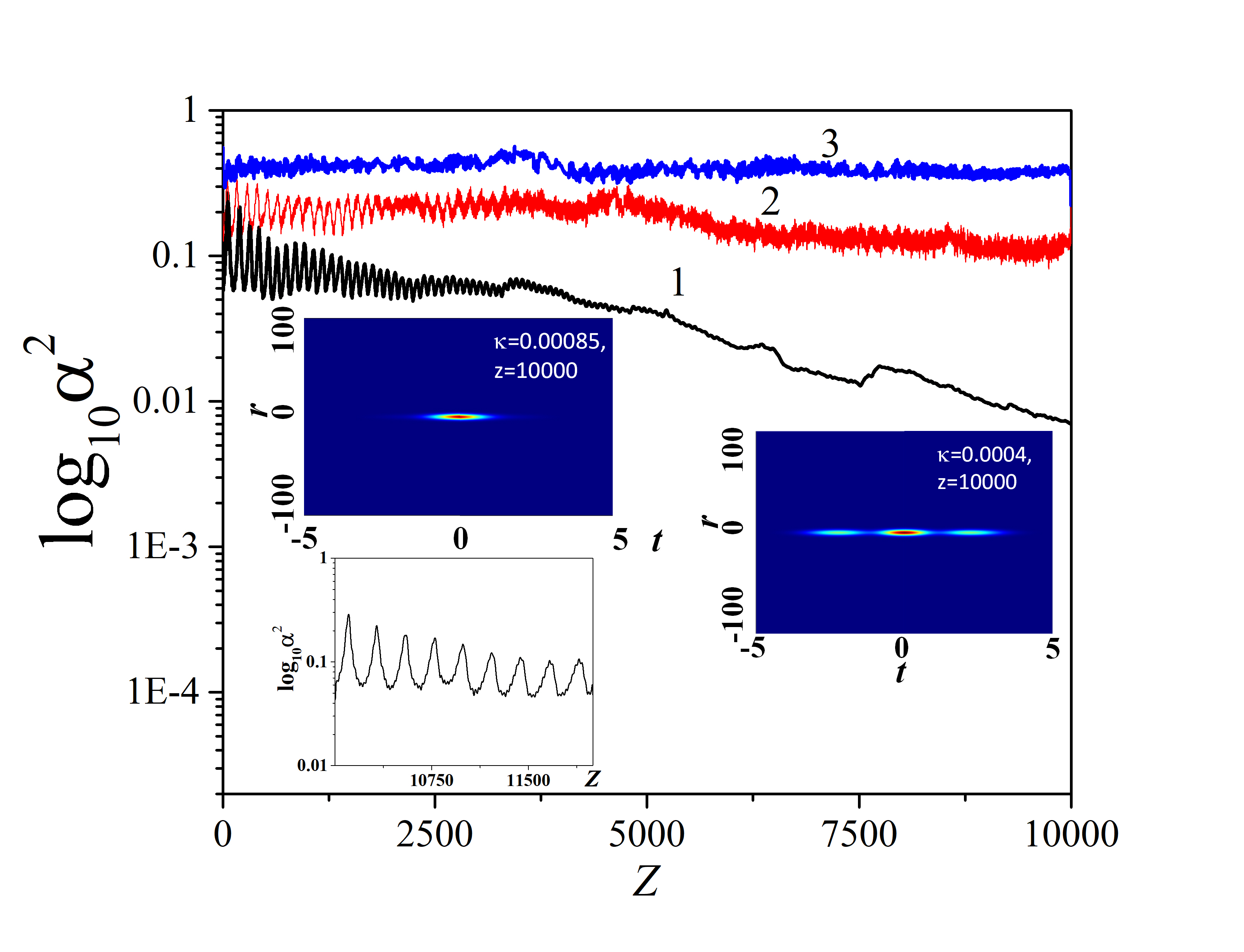}
\caption{\label{fig16} Averaged evolution of the maximum power for $\tau=$0.01, $\Lambda=-$0.001, $\nu=$0.01, and $\kappa=$0.00085 (1), 0.0007 (2), 0.0004 (3). The  insets show the contour plots of the LBs power at $Z$ shown in the insets. $const=0.3$ in Eq. (B1) (see Appendix). The left-bottom inset shows the oscillatory convergence of the solution (in logarithmic scale) within the $Z-$range of [10000,12000] for $\kappa=$0.00085.}
\end{figure}

Temporal confinement (i.e., $\nu>0$) reduces the ST DS temporal duration, in agreement with analytical predictions (see Fig. \ref{fig2} (b), and insets in Fig. \ref{fig16}). This effect is physically understandable as follows. Namely, the additional phase shift
$\nu t^2$ introduces a chirp on the propagating pulse. It shifts spectral components on the wings of the DS pulse, which then become closer to the edges of the parabolic gain-band. That increases losses on the temporal wings of the DS, and causes its temporal narrowing. At the same time, phase modulation may even suppress the DS for large values of $\tau$, i.e., in the case of a narrow gain bandwidth (see Table \ref{tab:table1}). Our numerical simulations confirm this conclusion. Moreover, numerical simulations demonstrate that the excitation of higher-order spatial modes occurs in parallel with temporal LB compression induced by a decrease of $\kappa$, without necessarily destroying the LB integrity and stability (see the curves and right inset in Fig. \ref{fig16}).
Therefore, the primary outcome of numerical simulations is that phase confinement in the time domain could lead to a stable LB generation.

\subsection{Beyond the radial symmetry approximation}

The radial symmetry condition which underlies Eq.(\ref{eq:1}) is a very strong assumption. Indeed, as it is well-known, multidimensional solitons suffer from numerous sources of perturbation, which lead to a rich zoo of different regimes, including those which destroy a soliton (e.g., see \cite{Malomed1,kartashov,malomed2016multidimensional,leblond2009stable,Kartashov11}). Hence, in order to prove soliton stability, full (3+1)-dimensional simulations are eventually required. In this case, the radially symmetric Lagrangian in Eq. (\ref{eq:1}) should be replaced by the Cartesian Lagrangian: ${{{\partial ^2}} \mathord{\left/
 {\vphantom {{{\partial ^2}} {\partial {x^2}}}} \right.
 \kern-\nulldelimiterspace} {\partial {x^2}}} + {{{\partial ^2}} \mathord{\left/
 {\vphantom {{{\partial ^2}} {\partial {y^2}}}} \right.
 \kern-\nulldelimiterspace} {\partial {y^2}}}$. 
 
\begin{figure}[htpb]
\includegraphics[width=7.5cm]{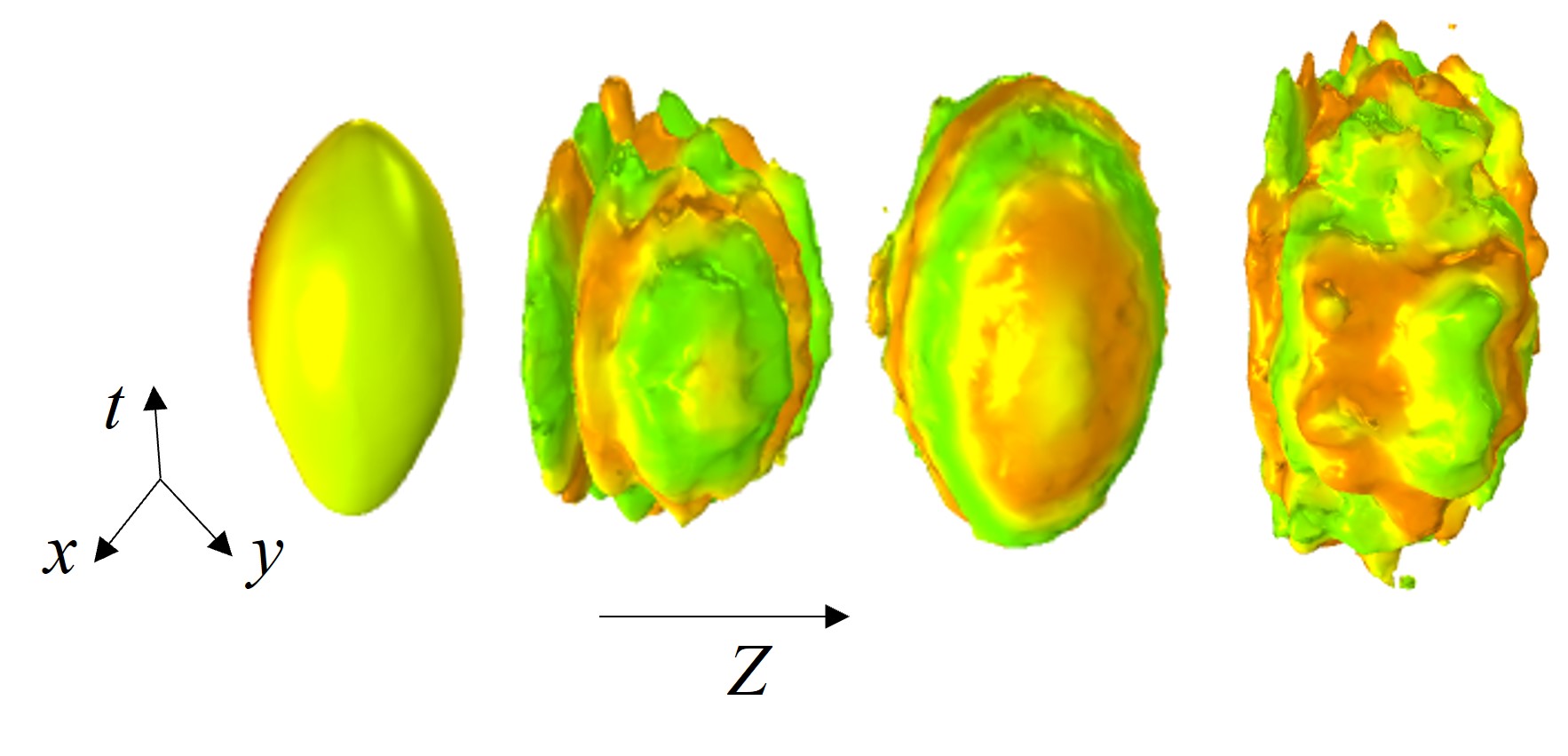}
\caption{\label{fig17} Four isosurfaces of a LB at different $Z$ for for $\kappa$=0.00085, $\Lambda$=-0.001, $\tau$=0.01, $const=1$ and $\nu$=0 in Eq. (B1) (see Appendix).}
\end{figure}

A comprehensive numerical analysis across the entire parameter space is beyond the scope of this work. Therefore, we limit ourselves to considering a couple of illustrative examples.
Fig. \ref{fig17} shows four shots of a LB, taken at different propagation distances $Z$ for $\nu$=0. Our simulations demonstrate that:
i) the LB preserves its integrity in the time dimension ($t$),
ii) spatial splitting is possible at some propagation distances,
iii) there are substantial LB profile distortions that violate the radial symmetry condition, and iv) the evolution is recurrent.

On the other hand, Fig. \ref{fig18} shows that, in the presence of a $t-$ confinement, these perturbations are wiped off, and the LB remains radially symmetric. Nevertheless, at some $Z$, the LB splits spatially in one direction, with a radial symmetry-breaking. Still, the LB profile remains smooth and radially symmetric during most part of the propagation distance. Significant LB perturbations may occur, depending on the initial conditions (see Fig. \ref{fig19}), but eventually, the LB preserves its shape.

\begin{figure}[htpb]
\includegraphics[width=7.5cm]{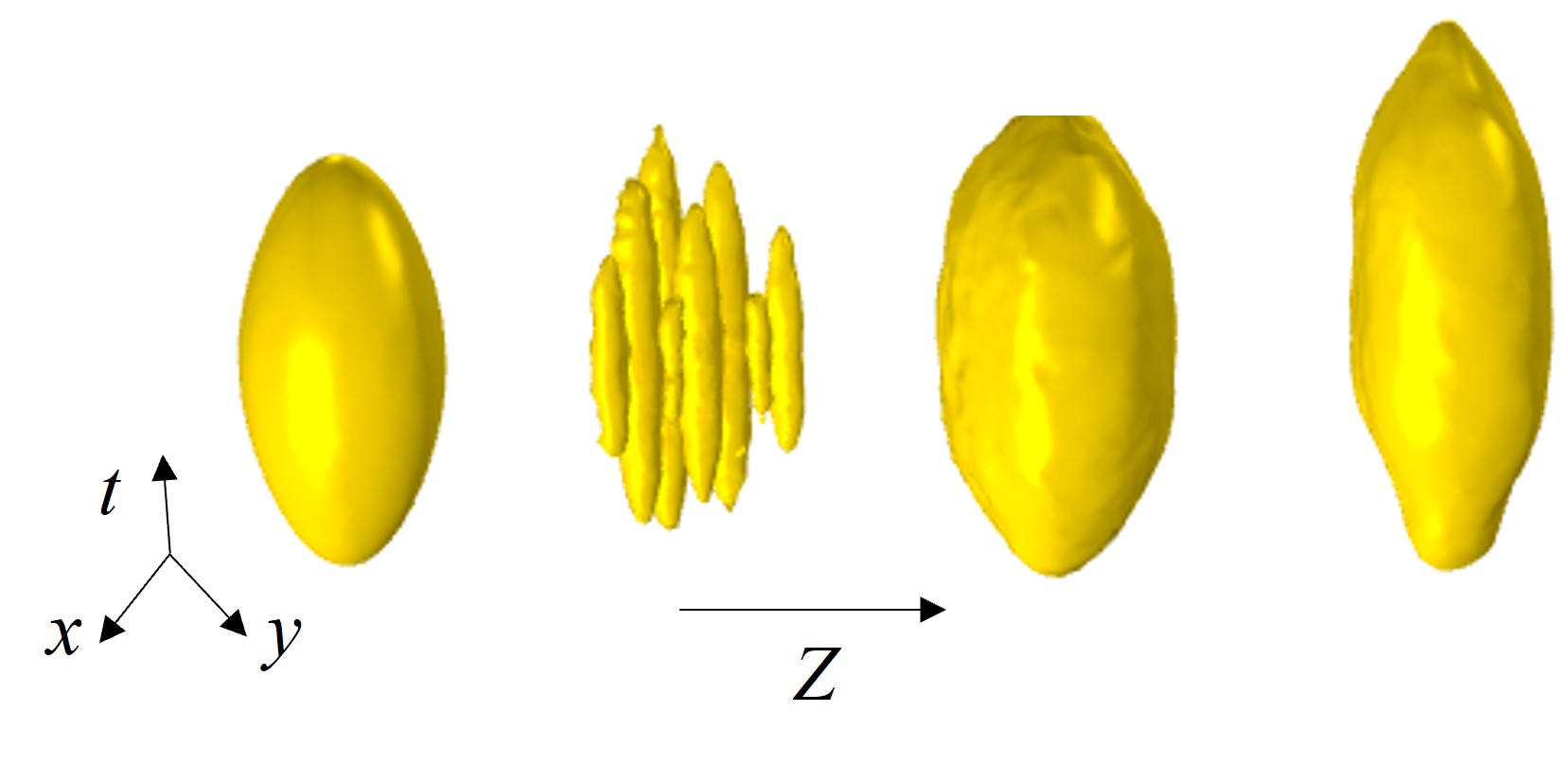}
\caption{\label{fig18} Four isosurfaces of a LB at different $Z$ for $\kappa$=0.00085, $\Lambda$=-0.001, $\tau$=0.01, $\nu$=0.01 and $const=1$ in Eq. (B1) (see Appendix).}
\end{figure}

\begin{figure}[htpb]
\includegraphics[width=6cm]{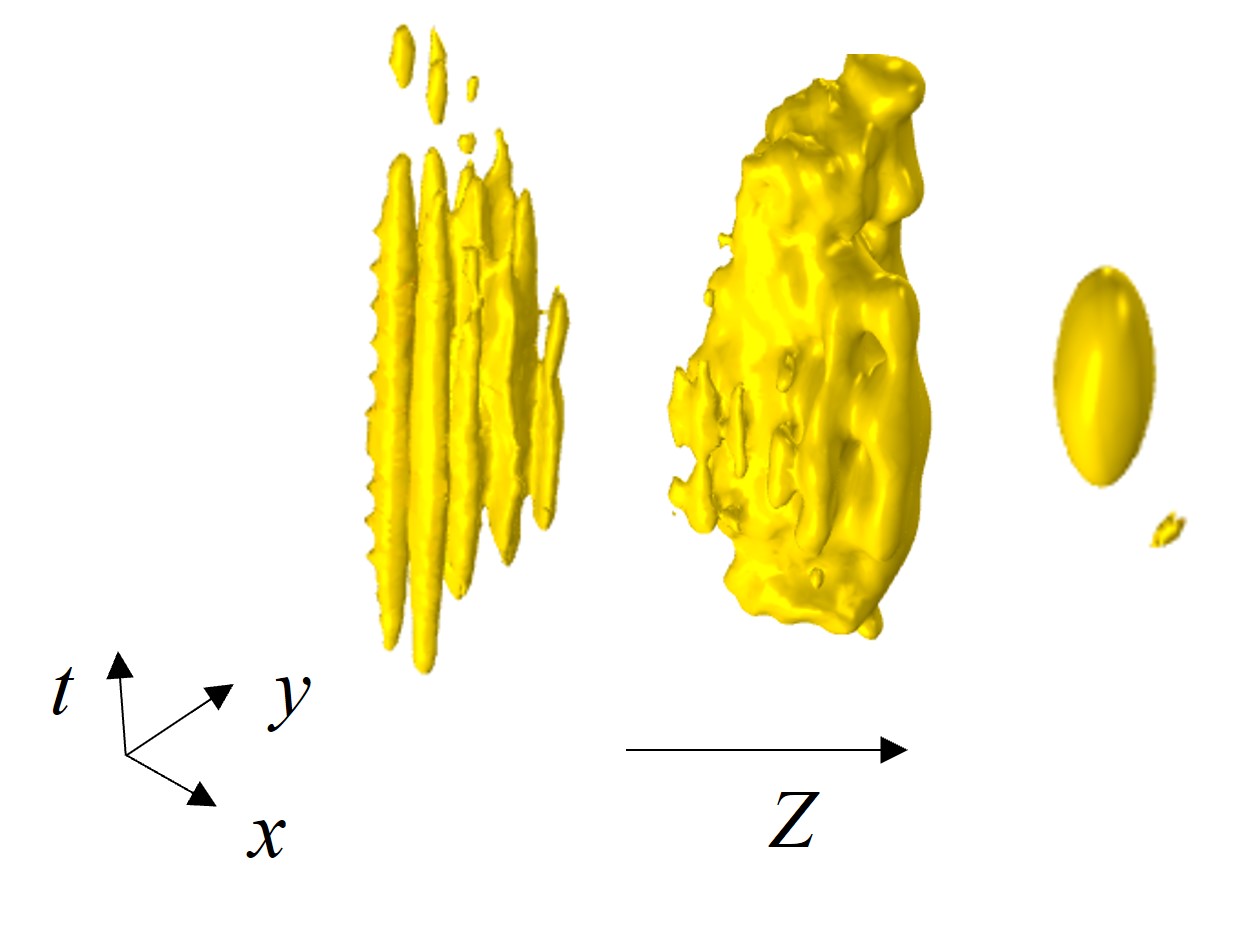}
\caption{\label{fig19} Four isosurfaces of a LB at different $Z$ for $\kappa$=0.00085, $\Lambda$=-0.001, $\tau$=0.01, $\nu$=0.01 and $const=0.3$ in Eq. (B1) (see Appendix).}
\end{figure}

\section{Discussion}

The comparison of the results of the variational approach and the numerical simulations reveals two important facts. Firstly, the variational analysis predicts a decrease in the LB energy as the dissipation gradient parameter $\kappa$ grows (see Fig. \ref{fig3}). This results from the degradation of the DS peak intensity. Such a result agrees with numerical results (see Fig. \ref{fig16}). However, the numerical simulations (curve 3, and corresponding inset in Fig. \ref{fig16}) demonstrate the existence of an ST DS in a ``forbidden region'' of the $\kappa-$parameter (as far as the variational analysis prediction is concerned). The ST DS tends to a stable intensity level after an initial transient process.

Secondly, numerical simulations demonstrate that an LB, which is formed at $\nu=0$, is subject to a multipulse generation instability (see Figs. \ref{fig12}-\ref{fig15}), which cannot be simply described in terms of the fundamental ansatz (2). Our outcome is that the resulting fragmented structure can be ``'fused'' by adding a ``temporal confinement'' mechanism, i.e., by a phase modulation with amplitude $\nu>0$. Such confinement leads to a stable LB, which is compressed in the time domain but may exhibit a ring-like structure in the spatial domain.

Simulations based on the (3+1)-dimensional model reveal the presence of radial symmetry-breaking, spatial splitting, and LB shape perturbations. Nevertheless, the LB preserves its integrity during most of the propagation distance. Thus, the presence of confinement in the $t-$coordinate suppresses LB perturbations, while preserving its radial symmetry.

\section{Conclusions}

Our work considers spatiotemporal dissipative solitons (LBs) with confinement in both spatial and temporal dimensions. Such a pancake-like potential could be a test-bed for stabilizing the ST SD in a low-dissipative Bose-Einstein condensate and graded-index multimode fiber lasers.

We based our analysis on the (2+1) and (3+1)-dimensional dissipative Gross-Pitaevskii equation, which was solved by both analytical and numerical approaches. In the first case, we used the variational approximation with a Gaussian-mode soliton ansatz. A direct numerical simulation based on the finite-element method was performed in the second case. The results obtained in the (2+1)-dimensional model framework are tested by (3+1)-dimensional simulations.

In terms of photonics, we found that some minimal curvature of the graded dissipative potential (the $\kappa-$ parameter in Eq. (\ref{eq:1})) is required to stabilize an ST soliton (LB). A 3D confinement due to phase modulation along a propagation axis was found to suppress the multipulsing effects and refines the LB spatio-temporal structure.

 A $t-$dependent phase-modulation in a fiber laser leads to confinement in the temporal dimension and relaxes the dependence of the light bullet stability on the graded dissipative potential. That means that an LB could be stabilized with weaker graded dissipation and exhibit significant temporal compression. However, the spatial structure of a light bullet for a low grading of the dissipative potential may acquire a complex multimode pattern.

Our findings suggest the following roadmap for controlling the ST DSs: i) A complex 3D confinement potential could be used, which involves a spatio-temporal localization of the gain. This can be obtained by combining transverse spatially graded dissipation with temporal phase modulation in a multimode fiber laser cavity (or loop); ii) The multiscale nature of the LB dynamics, involving both ``fast'' ($t-$) and ``slow'' ($Z-$) coordinates, should be exploited. In general, all parameters in Eq. (\ref{eq:1}) could become ($Z-t$)- dependent. Indeed, it was shown that the strategy of parameter management could stabilize spatiotemporal solitons \cite{Mayteevarunyoo_2020}; iii) It should be noted that the numerical aperture, connected with a wave-front curvature, can be significant in multimode fiber lasers. In this case, the paraxial approximation underlying the derivation of Eq. (\ref{eq:1}) may be invalid, and the generalized Helmholtz equation \cite{rol2022nonlinearity} should be used for the description of the transverse dynamics of MMF lasers and micro-waveguides.

\begin{acknowledgments}
This work has received funding from the European Union Horizon 2020 research and innovation program under the Marie Sk{\l}odowska-Curie grant No. 713694 (MULTIPLY) and the ERC Advanced Grant No. 740355 (STEMS). V.L.K. acknowledges Norwegian Research Council projects UNLOCK No. 303347 and MIR No. 326503.
\end{acknowledgments}

\appendix

\section{Evolution of DS parameters}

The variational approximation based on Eqs. (2--6) results in the following ordinary differential equations for the evolving parameters of the DS:
\begin{gather}
\frac{{d\theta }}{{dZ}} = \frac{1}{6} \left(3 +\frac{\alpha (Z)^2}{\rho (Z)^2}+12 \theta (Z)^2-\frac{3}{\rho (Z)^4}\right),\\
\frac{{d\psi }}{{dZ}} =\frac{3 \alpha (Z)^2-4 \left(3+\pi ^2\right) \tau  \psi (Z)}{3 \pi ^2 \upsilon(Z)^2}-\frac{2}{\pi ^2 \upsilon(Z)^4}+\nonumber \\ +\nu +2  \psi (Z)^2,\\
\frac{{d\alpha }}{{dZ}} =\frac{1}{15} \alpha (Z) \left(3 \pi ^2 \upsilon(Z)^2 \psi (Z)^2-\frac{5 \left(12+\pi ^2\right) \tau }{\pi ^2 \upsilon(Z)^2}\right)\nonumber+ \\+\frac{1}{15} \alpha (Z) \left(15 (\psi (Z)+2 \theta (Z)-\Lambda\right),\\
\frac{{d\upsilon }}{{dZ}} =\frac{8 \tau }{\pi ^2 \upsilon(Z)}-2 \upsilon(Z) \psi (Z)-\frac{16}{15} \pi ^2 \tau  \upsilon(Z)^3 \psi (Z)^2,\\
\frac{{d\rho }}{{dZ}} = -\rho (Z) \left(2 \theta (Z)+\kappa  \rho (Z)^2\right).
\end{gather}

\noindent The physical steady-state solutions of this system are expressed by Eqs. (7--8) \cite{nb}.

\section{Initial conditions for FEM}

As the initial conditions for the FEM simulations, we used the scalable expression for a DS amplitude,  the soliton relation between temporal width and amplitude, and the LB relation between beam width and amplitude \cite{raghavan2000spatiotemporal,nb}:

\begin{gather}
{\alpha _0} = const\sqrt {\frac{{3\left( {{\kappa ^2} - {\Lambda ^2} - {\Lambda ^4}} \right)}}{{\kappa \Lambda }}} ,\\
{\upsilon _0} = {{\sqrt 2 } \mathord{\left/
 {\vphantom {{\sqrt 2 } {{\alpha _0}}}} \right.
 \kern-\nulldelimiterspace} {{\alpha _0}}},\\
{\rho _0} = \frac{1}{6}\left( {\sqrt {\alpha _0^4 + 36}  - \alpha _0^2} \right).
\end{gather}

The initial spatial and temporal chirps (i.e., $\theta$ and $\psi$) were set as equal to zero.

\section{Parameters of effective dissipative nonlinearity}

It is challenging to connect ``ab ovo'' the nonlinear gain parameters (or self-amplitude modulation) of the reduced (1+1)-dimensional system (e.g., for a single-mode fiber laser) with those of the (2+1)-dimensional model (which describes an MMF laser). Nevertheless, some rough estimations could be analytically obtained \cite{haus1992analytic,lin1994analytical,yefet2013review}. Let us consider a Gaussian beam propagating in a Kerr-nonlinear medium \cite{nb2}. Its evolution can be estimated using a free-propagation model \footnote{Taking into account the graded refractive index requires more careful considerations.} by rescaling the imaginary part of the $q^{-1}$-parameter of the ABCD matrix:

\begin{equation}
    \frac{1}{{{q_0}}} =  - i\frac{{\sqrt {1 - K} \lambda }}{{\pi w_0^2}},
\end{equation}
\noindent where $w_0$ is the waist size, and $K=P/P_{cr}$ is the ratio of the beam power $P$ to the self-focusing critical power $P_{cr}$. Propagation over distance $z$ gives a new $q-$parameter: $q=(q_0+z)$, which needs proper rescaling, resulting in:

\begin{equation}
\frac{1}{q} = \frac{z}{{{z^2} + \frac{{{\pi ^2}w_0^4}}{{\left( {1 - K} \right){\lambda ^2}}}}} + \frac{{i\pi w_0^2}}{{\lambda \left( {1 - K} \right)\left( {{z^2} + \frac{{{\pi ^2}w_0^4}}{{\left( {1 - K} \right){\lambda ^2}}}} \right)}}.
\end{equation}

Then, the new imaginary part of the $q^{-1}$-parameter permits finding the new squared beam size $w$:

\begin{equation}
    \begin{array}{l}
\frac{{\pi {w^2}}}{\lambda } = \frac{{\left( {1 - K} \right)\lambda {z^2}}}{{\pi w_0^2}} + \frac{{\pi w_0^2}}{\lambda },\\
{w^2} = \frac{{\left( {1 - K} \right){\lambda ^2}{z^2}}}{{{\pi ^2}w_0^2}} + w_0^2.
\end{array}
\end{equation}

The following rough approximation permits to reduce the effect of axially distributed and radially graded dissipation to the action of a Gaussian aperture with radius $D$, localized at $z$. The induced loss is $\exp{(-D^2/w^2)}$, where

\begin{equation}
    \frac{{{D^2}}}{{{w^2}}} = \frac{{{D^2}}}{{\frac{{\left( {1 - K} \right){\lambda ^2}{z^2}}}{{{\pi ^2}w_0^2}} + w_0^2}}.
\end{equation}

Expansion of denominator over $K$ up to first-order allows for obtaining the intensity independent and dependent parts, respectively. Next, the modulation depth and the saturation parameters read as

\begin{equation}
\begin{array}{l}
\mu  = \frac{{{\lambda ^2}{z^2}}}{{{\pi ^2}{D^2}w_0^2}},\,\,\,\\
\varsigma  = P_{cr}^{ - 1}{\left( {1 + \frac{{{\lambda ^2}{z^2}}}{{{\pi ^2}w_0^4}}} \right)^{ - 1}}.
\end{array}
\end{equation}

\bibliography{article}

\end{document}